\documentclass{emulateapj}

\usepackage{natbib}
\usepackage{color}
\definecolor{red}{rgb}{0.7,0.1,0.1}
\definecolor{blue}{rgb}{0.2,0.2,0.8}
\definecolor{green}{rgb}{0.1,0.6,0.1}
\definecolor{white}{rgb}{1.0,1.0,1.0}
\definecolor{black}{rgb}{0.0,0.0,0.0}

\newcommand{\kms}{\ensuremath{\rm{km\,s}^{-1}}}

\newcommand{\rv}{$RV$}
\newcommand{\lv}{$LV$}
\newcommand{\rl}{$RL$}

\shorttitle{SFI++ disk galaxy scaling relations}
\shortauthors{Saintonge \& Spekkens}

\begin{document}

\title{Disk Galaxy Scaling Relations in the SFI++: \\ Intrinsic Scatter and Applications}

\author{Am\'{e}lie Saintonge\altaffilmark{1,2}}
\affil{Max Planck Institut f\"{u}r extraterrestrische Physik, Giessenbachstrasse, D-85748 Garching, Germany}
\email{amelie@mpe.mpg.de}

\author{Kristine Spekkens}
\affil{Royal Military College of Canada, Kingston ON, K7K 7B4, Canada}
\email{kristine.spekkens@rmc.ca}

\altaffiltext{1}{Max Planck Institut f\"{u}r Astrophysik, Karl-Schwarzschildstrasse 1, D-85748 Garching, Germany}
\altaffiltext{2}{Institute for Theoretical Physics, University of Zurich, CH-8057 Zurich, Switzerland}

\begin{abstract}
We study the scaling relations between the luminosities, sizes, and rotation velocities of disk galaxies in the SFI++, with a focus on the size-luminosity (\rl) and size-rotation velocity (\rv) relations. Using isophotal radii instead of disk scale-lengths as a size indicator, we find relations that are significantly tighter than previously reported: the correlation coefficients of the template \rl\ and \rv\ relations are $r=0.97$ and $r=0.85$, which rival that of the more widely studied \lv\ (Tully-Fisher) relation.  The scatter in the SFI++ \rl\ relation is 2.5-4 times smaller than previously reported for various samples, which we attribute to the reliability of isophotal radii relative to disk scale-lengths.  After carefully accounting for all measurement errors, our scaling relation error budgets are consistent with a constant intrinsic scatter in the \lv\ and \rv\ relations for velocity widths $\log W \gtrsim 2.4$, with evidence for increasing intrinsic scatter below this threshold. The scatter in the \rl\ relation is consistent with constant intrinsic scatter that is biased by incompleteness at the low-$L$ end.   Possible applications of the unprecedentedly tight SFI++ \rv\ and \rl\ relations are investigated.  Just like the Tully-Fisher relation, the \rv\ relation can be used as a distance indicator: we derive distances to galaxies with primary Cepheid distances that are accurate to 25\%, and reverse the problem to measure a Hubble constant $H_0=72\pm7$ \kms\ Mpc$^{-1}$.  Combining the small intrinsic scatter of our \rl\ relation ($\epsilon_{int}=0.034\pm0.001 \log [h^{-1} \mathrm{kpc}]$) with a simple model for disk galaxy formation, we find an upper limit on the range of disk spin parameters that is a factor of $\sim 7$ smaller than that of the halo spin parameters predicted by cosmological simulations. This likely implies that the halos hosting Sc galaxies have a much narrower distribution of spin parameters than previously thought.
\end{abstract}


\section{Introduction}
\label{intro}

 The observed correlations between spiral galaxy luminosities ($L$), sizes ($R$) and rotation velocities ($V$) have long been exploited to provide insight into the large-scale galaxy distribution and the nature of disk galaxies themselves. Calibrated Tully-Fisher \citep{tf77} scaling relations built from the distance-dependent $L$ and the distance-independent $V$ (we henceforth refer to this as the \lv\ relation)  have been extensively used as distance indicators, while scatter in $L$ caused by deviations from the Universal expansion probe the large-scale mass distribution. The tightness of the \lv\ relation as well as that between $R$ and $V$ (the \rv\ relation) and between $R$ and $L$ (the \rl\ relation) over decades in mass also provide important clues to the process of galaxy formation. Most theories connect the structure of galactic disks to that of their parent dark matter halos \citep[e.g.][see below]{fall80}, themselves well-constrained by simulations of halo assembly. A careful consideration of the scaling relations in the context of galaxy formation thus probes how baryons collapse to form disks as well as the properties of the dark matter halos that host disk galaxies. 

The \lv\ relation has been used routinely over the last 30 years to compute distances to nearby galaxies and their peculiar velocities relative to the Hubble flow.  With typical accuracies of $15-20$\%, \lv\ distances have been used successfully for example to determine the 3D structure of the Virgo cluster  \citep[e.g.][]{tully84,pierce88,yasuda97,gavazzi99} and to compute the value of $H_0$, with accuracy comparable to cosmology experiment-derived values \citep[e.g.][hereafter M06]{giovanelli97Ho,sakai00,masters06}.  Because of the usefulness, accuracy and relative ease of measurement of \lv\ distances, large galaxy samples have been assembled for this purpose at most optical and near-infrared wavelengths \citep[e.g.][]{aaronson82,mathewson92,willick97,theureau98}. 

The SFI++ \citep[][hereafter S07]{springob07} is a sample of $\sim 5000$ local galaxies, one of the largest of its kind, and represents the culmination of 15 years worth of work by Giovanelli, Haynes and collaborators (hereafter collectively referred to as ``the Cornell group") towards measuring peculiar velocities using \lv\ distances.  The SFI++ builds on the earlier {\it Spiral Field I-band} \citep[SFI:][]{giovanelli94,giovanelli95,haynes99b,haynes99}, {\it Spiral Cluster I-band} \citep[SCI:][]{giovanelli97}, and {\it Spiral Cluster I-band 2} \citep[SC2:][]{dale99b,dale99a} samples. It also includes additional data from the previously unpublished {\it Spiral Field I-band 2} sample (SF2) and from the southern hemisphere surveys of \citet{mathewson92} and \citet{mathewson96}, with photometry re-processed to match that of the other catalogs.  The sample contains both field and cluster galaxies, mostly of type Sc. A subset of the cluster galaxies comprise a {\it template} sample used to calibrate the \lv\ relation via the ``basket-of-clusters" technique \citep[][M06]{giovanelli97}. The remaining catalog members comprise the {\it nontemplate} sample, from which peculiar velocities are estimated by application of the template \lv\ relation (S07). 

The \lv, \rv, and \rl\ scaling relations are also important diagnostics of galaxy formation: $L$ is fundamentally a tracer of stellar mass, $R$ a tracer of disk specific angular momentum, and $V$ a tracer of (dark + luminous) mass. In the standard model for galaxy formation, disks form through the dissipative collapse of hot gas within the potential wells of their dark matter halos, conserving specific angular momentum \citep{white78,fall80,dalcanton97,mo98}.  This scenario relates $L$, $R$ and $V$ of galactic disks directly to the virial properties of their parent halos, which can be reliably measured from collisionless simulations \citep[e.g.][]{maccio08}. These models have been extensively applied to the \lv\ relations originally derived as distance indicators \citep[e.g.][]{mo98,dalcanton97,vdb98,somerville99,vdb00,navarro00,firmani00} and to the \rl\ relation to a lesser extent \citep{salpeter96,dejong00,graham02,shen03,avila08}. They can typically be ``tuned" to match the slopes, zero-points and scatters of a given relation.  However, simultaneously reproducing all scaling relations and the galaxy luminosity function is a notoriously difficult problem, as it involves both cosmological initial conditions and the physics of star formation and feedback \citep[e.g.]{vdb00,somerville99,bell03,croton06,dutton07}.

In light of their importance in constraining galaxy formation models, scaling relations are now being constructed specifically for this application \citep[e.g.][]{kannappan02,kauffmann03b,shen03,zavala03,pizagno05,courteau07,pizagno07}. Of most relevance to the present study, \citet[][hereafter C07]{courteau07} and \citet[][hereafter P05]{pizagno05} have constructed the $LRV$ scaling relations from photometry and kinematics comparable to that in the SFI++.  C07 combines subsets of four extant \lv\ distance samples into a catalog of 1300 galaxies with $V$ estimated from optical rotation curves, $R$ and $L$ derived from $I$-band scale-lengths and apparent magnitudes respectively, as well as as an estimate of $R$ from 2MASS $K$-band effective radii. The raw magnitudes, velocity widths and scale lengths are inhomogeneous, but C07 apply the same inclination and extinction corrections to all galaxies. By contrast, P05 use a significantly smaller sample of 81 disk-dominated galaxies, but with homogeneously measured $V$ from optical rotation curves, $L$ from SDSS \citep{york00} $i$-band magnitudes, and $R$ from $i$-band scale-lengths via 2-dimensional bulge-disk decompositions.  Additionally, they use SDSS $g-r$ colors to estimate stellar masses. 
 
\citet{dutton07} and \citet{gnedin07} apply galaxy formation models to simultaneously fit the C07 and P05 $LRV$ relations, respectively. Both claim to match the scaling relations, their residuals and scatter as well as the galaxy luminosity function, but not without invoking non-standard parameters such as low mass-to-light ratios \citep{gnedin07} or processes such as halo expansion \citep{dutton07}. These studies have clearly demonstrated the need for large, homogeneous samples of galaxies from which scaling relations with well-understood residuals, scatters and uncertainties can be derived. 
 
 The exponential scale-length $r_d$ of galaxy disks is typically adopted as the measure of $R$ in disk galaxy scaling relations. However, disk scale lengths are notoriously difficult quantities to measure. For low-inclination systems ($i \leq 50^\circ$), tests of measured scale-length reliability typically return uncertainties of 10-20\%  \citep[e.g.][]{schombert87,byun95,dejong96,macarthur03,fathi10}, while comparisons of $r_d$ reported by different authors for the same galaxy reveal a scatter of $\sim25\%$ \citep{knapen91,mollenhoff04}.  Most scaling relation samples contain galaxies with $i \gtrsim 60^\circ$ to mitigate uncertainties in the inclination correction required to estimate $V$. In the optical, disks are partially opaque in the region where $r_d$ is measured \citep[e.g.][]{giovanelli94, giovanelli95}, and internal extinction is a significant additional source of uncertainty in the derivation of $R$ in the scaling relations. Recipes to correct $r_d$ for internal extinction exist \citep[e.g.][]{byun92,byun94,giovanelli94,giovanelli95,graham01,masters03,mollenhoff06,graham08}, and imply that large ($\gtrsim 20\%$), uncertain (by $\sim 30\%$) corrections are required even at moderate inclinations of $i \sim 70^\circ$. 
 
 Given the large and potentially systematic uncertainties in $r_d$, it is perhaps not surprising that the scaling relations constructed using this parameter  show weaker correlations and larger scatters than the \lv\ relation (e.g.\ C07, P05), and have been correspondingly less well-studied than the latter in both the distance indicator and galaxy formation contexts.  It therefore seems worthwhile to explore other measures of $R$ than $r_d$, such as isophotal radii, to construct the $LRV$ scaling relations: preliminary work by \citet{spekkens05} and \citet{saintonge08} has demonstrated promise in this approach.

 While originally designed to compute peculiar velcocities from \lv\ distances, the SFI++ can also be exploited to study the \rv\ and \rl\ scaling relations of Sc galaxies.  In addition to well-characterized measures of $L$ and $V$, a significant subset of SFI++ galaxies have homogeneously measured $r_d$ as well as the radius $r_{23.5}$  measured at the $\mu_I = 23.5$  mag arcsec$^{-2}$  isophote \citep[][S07]{haynes99}. Detailed studies using earlier compilations of the catalog explore and correct for internal extinction effects in these parameters \citep{giovanelli94,giovanelli95}.   Thus while the ideal approaches toward sample selection and parameter measurement may differ when compiling samples for \lv\ distances versus galaxy formation analyses \citep[][C07, see also \S\ref{applications}]{pizagno07,avila08}, the size, quality and homogeneity of the SFI++ are unrivalled for both applications. 
 
  In this paper, we present the \lv, \rv\ and \rl\ scaling relations, their residuals and their scatter for a subset of the SFI++ template and nontemplate samples. Since the \lv\ relation has already been extensively studied by M06, we focus here on the \rv\ and \rl\ relations. We argue that  $r_{23.5}$ is superior to $r_d$ as a measure of $R$, and present \rl\ relations with observed scatters that are factors of $\sim2.5-4$ smaller than previously found. We derive detailed error budgets for all relations, and estimate the contribution of measurement errors to the observed scatters. We then discuss the applications \rv\ and \rl\ relations for measuring redshift-independent distances and constraining galaxy formation models, respectively. For clarity, we present the mathematical details related to the derivation of the SFI++ scaling relation parameters, their uncertainties, and the scaling relation error budgets in a pair of appendices.  In all sections except \S\ref{applications}, we adopt a value of $H_0=100 h^{-1}$ \kms Mpc$^{-1}$ for distance-dependent quantities.

\section{Data}
\label{data}

We select our sample from the SFI++ (see \S\ref{intro}). Specifically, we include all SFI++ galaxies for which radii at the $\mu_I = 23.5\,$mag$\,$arcsec$^{-2}$ isophote and disk scale-lengths have been homogeneously measured from $I$-band photometry \citep[][S07]{haynes99}. 
We treat the template and nontemplate galaxies separately  because  (1) as discussed in \S\ref{budget}, peculiar velocities and incompleteness biases affect them differently and (2) different subsamples are appropriate for different applications.   Throughout this paper, we refer to the 664/807 SFI++ template galaxies and 3655/4054 SFI++ nontemplate galaxies that meet our selection criteria as the {\it template subsample} and {\it nontemplate subsample}, respectively, and perform our analysis separately on each.

\subsection{Measurements, Corrections and Error Estimates \label{meas_errors}}

Absolute $I$-band magnitudes and homogenized velocity widths for the SFI++ are presented in S07. The derivation of these quantities, which draws on the work of the Cornell group over the past 15 years \citep[][M06, S07]{giovanelli94,giovanelli95,giovanelli97,haynes99,catinella05,springob05,catinella07}, is also summarized in that paper. Because these derivations are paramount to a rigorous computation of the scatter in the scaling relations presented here, we compile the relevant measurement, correction and uncertainty estimate equations in Appendices~\ref{ap_mcor}~and~\ref{ap_wcor}.  We provide a brief description of these parameters and the adopted corrections in \S\ref{meas_mcor} and \S\ref{velwidths}.

A number of disk size measures are also available for the template and nontemplate subsamples. In \S\ref{meas_rcor}, we justify our choice of the radius at the $\mu_I = 23.5\,$mag$\,$arcsec$^{-2}$ isophote as the disk size and give an overview of its derivation. The related mathematical details are presented in Appendix~\ref{ap_rcor}. 

We follow the same prescription for computing distances to SFI++ galaxies as in S07, which we describe in \S\ref{distances}.

\subsubsection{Luminosities \label{meas_mcor}}

Luminosities in the SFI++ are expressed in terms of absolute $I$-band magnitudes, $M_I$. Their derivation is explained in S07:  apparent magnitudes extrapolated to 8 disk scale-lengths are extracted from $I$-band photometry \citep{haynes99}. They are corrected for Galactic extinction using the values of \citet{schlegel98} and internal extinction using the relations of \citet{giovanelli94, giovanelli95}. The type-dependent k-correction of \citet{han92} is also applied. Absolute magnitudes are then computed using measured galaxy or cluster redshifts as summarized in \S\ref{distances}. Uncertainties on $M_I$ are computed by propagating (uncorrelated) measurement errors on the apparent magnitude and Galactic and internal extinction corrections. The mathematical details of these computations, first presented in a series of papers by the Cornell group \citep[][M06, S07]{giovanelli97,haynes99}, are compiled in Appendix~\ref{ap_mcor}.

\subsubsection{Rotation Velocities \label{velwidths}}
Rotation velocities in the SFI++ are expressed in terms of the logarithm of the velocity width, $\log{W}$, and are presented in S07. As explained in that paper, $\log W$ is derived from either single-dish HI profiles or optical rotation curves (ORCs) as in \citet{springob05} or \citet{catinella05}, respectively. HI widths are corrected for instrumental broadening and turbulence, while ORC widths are measured from a parametric fit to the folded rotation curve and homogenized with the HI widths using the relations derived by \citet{catinella07}. All velocity widths are corrected for cosmological stretching, and for inclination using measured $I$-band ellipticities \citep[][S07]{haynes99, giovanelli97}. Uncertainties on $\log W$ are computed by propagating (uncorrelated) measurement errors on the velocity width and ellipticity, as well as on the intrinsic axial ratio of the disk. The mathematical details of these computations, first presented in a series of papers by the Cornell group \citep[][M06, S07]{giovanelli97,haynes99,springob05,catinella05,catinella07}, are compiled in Appendix~\ref{ap_wcor}.

\subsubsection{Sizes \label{meas_rcor}}

 As discussed in \S\ref{intro}, deprojected disk scale-lengths $r_d$ are difficult to reliably measure despite their widespread use to construct scaling relations: the value of $r_d$ extracted from the surface brightness profile depends on the subjective process of choosing the profile's exponential region (``marking the disk"), and extinction corrections applied to $r_d$ are large and uncertain \citep[e.g.][]{giovanelli94,giovanelli95}.  Given these effects, it is clear that the actual uncertainty in  extinction-corrected $r_d$ is much larger than the $\sim15\%$ measurement errors that are generally assigned \citep[e.g.][]{schombert87,byun95,dejong96,macarthur03,fathi10}. 
 
  We therefore adopt the isophotal radius $r_{23.5}$ measured at the $\mu_I = 23.5$ mag arcsec$^{-2}$  isophote as the SFI++ measure of disk size. This quantity is straightforward to measure homogeneously from the high-quality SFI++ photometry \citep[][see also \citealt{courteau96}]{haynes99}, and does not require marking the disk. It is measured at a location far enough from the bulge that the latter does not contribute significantly to the light. Its inclination dependence was studied by \citet{giovanelli94,giovanelli95}: they found that galaxies with $M_I > -21\,$mag are completely transparent at $\mu_I = 23.5$ mag arcsec$^{-2}$, making the extinction correction also straightforward.

 We define the isophotal radius that we adopt for the SFI++ disk size as $R_{23.5}$ in units of $h^{-1}\,\mathrm{kpc}$, and express it logarithmically. We present a detailed derivation of $\log R_{23.5}$  in Appendix~\ref{ap_rcor}. Briefly, the radius corresponding to the $\mu_I = 23.5$ mag arcsec$^{-2}$ isophote is measured from $I$-band photometry as described by \citet{haynes99}. The same Galactic extinction and k-corrections adopted to compute $M_I$ (\S\ref{meas_mcor}) are then applied, and the resulting values are corrected for cosmological surface brightness dimming and cosmological stretching. We then correct for internal extinction using the relations derived by \citet{giovanelli95}, 
 and convert from angular to physical sizes using the measured galaxy or cluster redshifts as explained in \S\ref{distances}.

Following the approach of \citet{giovanelli97}, we derive uncertainties for each value of $\log R_{23.5}$ by propagating uncorrelated measurement uncertainties. Specifically, we include uncertainties on the raw isophotal radii, disk scale-lengths, Galactic and internal extinction correction, as well as the measured uncertainties on the disk ellipticity. 
 
\begin{figure}
\plotone{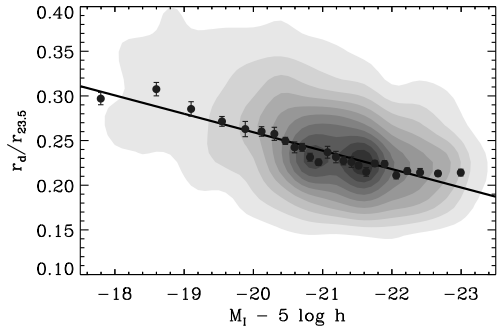}
\caption{Ratio of the measured disk scale-length $r_d$ to the isophotal radius $r_{23.5}$ (computed using eq.~\ref{r235}) as a function of $M_I$ for the nontemplate subsample. The points show the median ratio in equally populated absolute magnitude bins, and their errorbars represent the uncertainties on the position of the median determined by bootstrapping. The solid line shows the best linear fit to the points.
\label{rd_r235}}
\end{figure}

 Fig.~\ref{rd_r235} shows the ratio of $r_d$ to $r_{23.5}$ (the latter given by eq.~\ref{r235}) as a function of $M_I$ for the nontemplate subsample. As explained by \citet{giovanelli95}, the dependence of $r_d/r_{23.5}$ on $M_I$ stems from the systematic change of disk central surface brightnesses with this quantity. The solid line in Fig.~\ref{rd_r235} is the best linear fit to the median values indicated by the points:
 \begin{equation}
  \langle \frac{r_d}{r_{23.5}} \rangle = 0.672+0.0206 (M_I - 5 \log h) \;.
  \label{rdeq}
 \end{equation}
 We multiply $R_{23.5}$ by this relation to convert our disk sizes to scale-length units when comparing with previous studies (\S\ref{scalingrelations}) and galaxy formation models (\S\ref{gal_form}).  

\subsubsection{Distances \label{distances}}

Following M06 and S07, we adopt different values of the CMB-frame redshift $cz$ in the computation of $M_I$ and $\log R_{23.5}$ for the template and nontemplate subsamples.  For the template subsample, ``in" galaxies are assigned $cz$ of their parent cluster measured by M06, while distances to ``in+" galaxies are computed from $cz$ measured for each galaxy assuming pure Hubble flow (see M06 for the definition of ``in" and ``in+").  For the nontemplate subsample, distances are computed from $cz$ for each galaxy, assuming pure Hubble flow. 

 Note that S07 compute peculiar velocities for nontemplate SFI++ galaxies, which they take as the offset between $M_I$ and the template \lv\ relation from M06. However, only part of the scatter in this relation stems from peculiar velocities (see \S\ref{budget}), which implies that the values computed by S07 are upper limits. We therefore assume pure Hubble flow for our nontemplate subsample galaxies, and treat peculiar velocities as a source of intrinsic scatter in the relations.

\section{Scaling Relations in the SFI++ }
\label{scalingrelations}

 We now construct \lv, \rv\ and \rl\ scaling relations for SFI++ template and nontemplate subsamples. 

\subsection{The \lv, \rv, and \rl\ Relations\label{vrl}}

\begin{figure*}
\epsscale{1.0}
\plotone{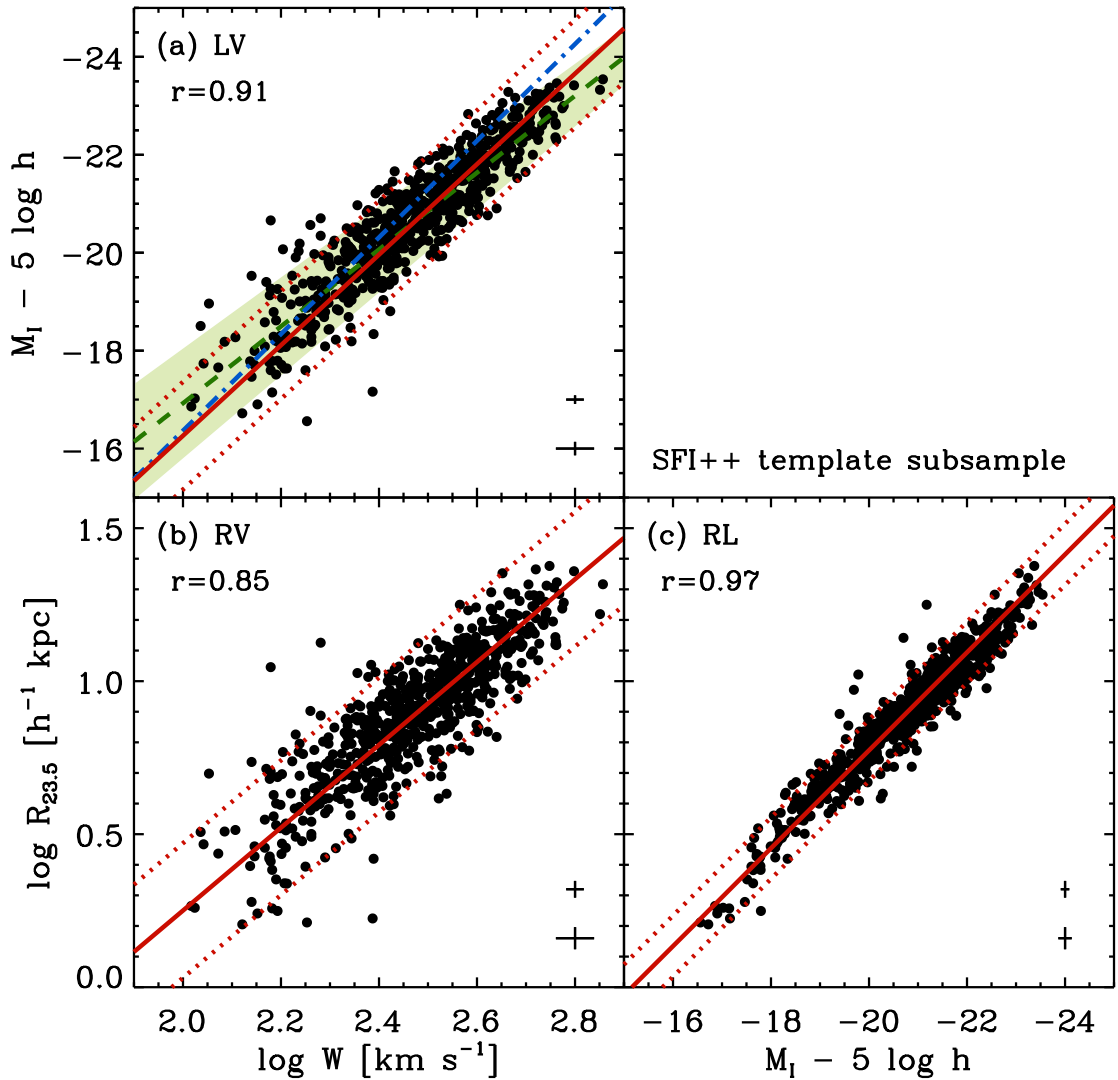}
\caption{Scaling relations for the SFI++ template subsample: $(a)$ the luminosity-velocity (\lv) relation, $(b)$ the size-velocity (\rv) relation and $(c)$ the size-luminosity (\rl) relation. The morphological corrections of eqs.~\ref{deltaLV},~\ref{deltaRL}~and~\ref{deltaRV} have been applied in panels $(a)$, $(b)$ and $(c)$ respectively. In each panel the best orthogonal linear fit is plotted as a solid line, and the $2\sigma$ scatter $2\epsilon_{obs}$ in the $y$-axis variable about that fit is delimited by dotted lines.  The Pearson correlation coefficient of each relation is given in the top-left corner of each panel. The crosses in the bottom-right corner of each panel show the median measurement uncertainty in the sample (top) and the the median uncertainty $\pm$ 3 times the median absolute deviation of this quantity. In $(a)$, the dashed line and shaded region shows the SFI++ template \lv\ relation derived by M06 and the dot-dashed line the relation of \citet{sakai00}. A color version of this figure is available in the electronic edition of the Journal.
\label{VRL_T}}
\end{figure*}

\begin{figure*}
\epsscale{1.0}
\plotone{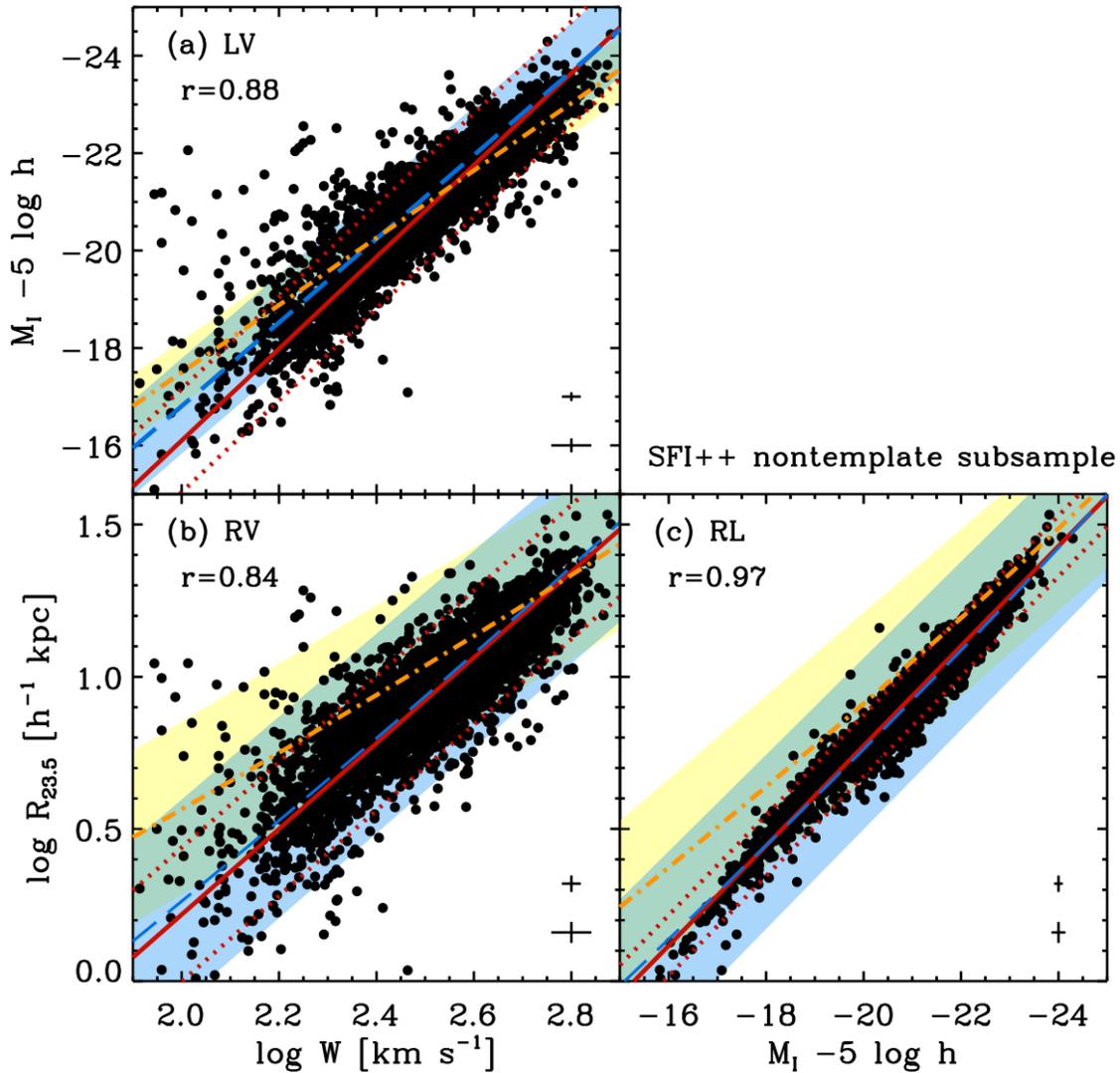}
\caption{Same as Fig.~\ref{VRL_T}, but for the SFI++ nontemplate subsample. The long-dashed lines and light (blue) shaded regions show the best-fitting relations and ($2\sigma$) scatter determined by C07,  and the dash-dotted lines and dark (yellow) shaded regions show the best-fitting relations and scatter determined by P05. The parameters in these studies are converted to our notation as described in the text. A color version of this figure is available in the electronic edition of the Journal.
 \label{VRL_NT}}
\end{figure*}

Figures~\ref{VRL_T}~and~\ref{VRL_NT} show the \lv, \rv, and \rl\ relations for the template and nontemplate subsamples defined in \S\ref{data}, respectively. All scaling relations were modeled using an orthogonal linear fitting method which takes into account measurements on $x$ and $y$ simultaneously.  A reliable estimate of the scatter $\epsilon_{obs}$ in the $y$-axis variable about the best-fitting linear relation is computed by applying Tukey's biweight to derive a robust measurement of the dispersion\footnote{In the case of an outlier-free, Gaussian distribution this approach reduces exactly to the classical standard deviation, but in the presence of strong outliers it provides much more stable results.}.  The best-fitting orthogonal relations and $2\epsilon_{obs}$ intervals are plotted as solid and dotted lines in each panel of the figures, and the corresponding fit parameters are given in Table~\ref{templates}.  

The morphological corrections in eqs.~\ref{deltaRL},~\ref{deltaRV}~and~\ref{deltaLV} have been applied to all of the relations in Figs.~\ref{VRL_T}~and~\ref{VRL_NT} (see \S\ref{morphology}). They are therefore {\it representative of Sc galaxies specifically}. We present the best-fitting linear relations to both the uncorrected and corrected relations in Table~\ref{templates}.  As a simple measure of the tightness of the relations, the Pearson correlation coefficient $r$ is computed in each case and is also given in Table~\ref{templates}, as well as in the  top-right corner of each panel in Figs.~\ref{VRL_T}~and~\ref{VRL_NT}.

The top cross in the bottom-right corner of each panel in Figs.~\ref{VRL_T}~and~\ref{VRL_NT} shows the median measurement uncertainty $\epsilon_{mes}$ in the sample, and the cross below it shows the median $\pm$ 3 times the median absolute deviation of that distribution. 
We discuss the sources of scaling relation scatter in detail in \S\ref{budget}; for now, we point out that, as found in many previous studies, $\epsilon_{obs}$ easily exceeds the median $\epsilon_{mes}$.

There is a tight linear correlation between $L$ and $V$ in the SFI++; we refer the reader to M06 for an exhaustive discussion of the SFI++ \lv\ relation. For the entire SFI++ template sample, M06 derived an incompleteness-corrected, morphology-corrected template \lv\ relation parametrized by $M_{I} -5\log h=-20.85-7.85(\log W -2.5)$. This relation and its median scatter are shown in Fig.~\ref{VRL_T}a as the dashed line and shaded region. The M06 relation is slightly shallower than the one derived here due to small differences in sample composition: our template subsample contains only $\sim 80\%$ of the galaxies used by M06 (see \S\ref{data}).

We note that M06 correct the SFI++ \lv\ template for incompleteness, an effect that biases its slope low and its zero point high because we are more likely to detect the brighter galaxies near the flux limit of a survey (thus at low $V$). We do not correct the template subsample used here for incompleteness. As a cursory investigation of the impact of incompleteness on our sample, we fit the \lv\ relation only to galaxies with $\log W > 2.4$, which M06 demonstrate are unaffected by incompleteness in the SFI++ template (see also \S\ref{incompleteness}). As expected, the re-fitted relation has a steeper slope and brighter zero point, pulling it farther away from the M06 result shown in Fig.~\ref{VRL_T}a. Nonetheless, the slope and zero point of this re-fitted \lv\ relation are consistent with those in Table~\ref{templates} within their errors. At least for the \lv\ relation, incompleteness therefore does not produce a strong bias relative to the uncertainties in our linear fits, in part because of the smaller sample adopted here compared to M06 (and thus the larger statistical uncertainties). We discuss the effect of incompleteness on the scaling relation scatter in \S\ref{budget}.

Taking similarities in sample selection, photometric band and parameter measurement into account, it is most straightforward to compare our nontemplate subsample relations with the results of P05 and C07.  The best-fitting relations and $2\sigma$ scatters from those studies are overplotted on our nontemplate relations in Fig.~\ref{VRL_NT}. To carry out the comparison, luminosities are converted to magnitudes assuming $M_{\odot,I}=4.19$, and we correct for the slight difference in photometric band of the P05 analysis using $\log L_I = \log L_i + 0.036$ (C07). We change from rotation velocities to velocity widths by multiplying by two. Finally, we convert from the scale-length units used in P05 and C07 to the isophotal radius units adopted here using the relation in eq.~\ref{rd_r235}. 
Considering the differences in sample selection and parameter estimation, there is reasonable agreement between the best-fitting \lv\ relation and scatter found here and those of P05 and C07 (dash-dotted and long-dashed lines in Fig.~\ref{VRL_NT}, respectively). 

We also find strong correlations between $R$ and both $V$ and $L$ in the SFI++. Fig.~\ref{VRL_NT} shows that when converted to our units, the best-fitting C07 relations agree very well with ours. This is expected given the significant overlap between the samples, but reassuring to confirm because of the different parameters  and corrections adopted in both studies. By extension from the discussion in C07 and more recent studies, there is broad general agreement between our best-fitting relations and others in the literature \citep{dejong00,shen03,graham08,fathi10}.  As suggested by \citet{dutton07}, the systematically larger scale-lengths found by P05 may stem from differences in galaxy selection and parameter derivations relative to other studies, and is likely exacerbated by their small sample.

In contrast to the similarity between our best-fitting \rv\ and \rl\ slopes and zero points and those in the literature, the high correlation coefficients of our relations (Table~\ref{templates}) provide the first indication that they are significantly tighter than has been previously reported: for example, we measure $r=0.84$ and $r=0.97$ for the \rv\ and \lv\ relations in the nontemplate subsample, while C07 measure an average of $r\sim0.65$ for the analogous relations. 

Fig.~\ref{VRL_NT} illustrates how the scatter in our best-fitting \rv\ and \rl\ relations (particularly the latter) is significantly smaller than that reported by P05 and C07. Focussing on the more widely studied \rl\ relation, we find an average scatter of $\epsilon_{obs}=0.05 \log (h^{-1} \mathrm{kpc})$ for both the template and nontemplate subsamples (Table~\ref{templates}). Converted to our units,
C07 report a scatter of $\sigma_{logR} \sim 0.14 \log (h^{-1} \mathrm{kpc})$ using a sample that is similar to and overlaps with the SFI++ in many respects. P05 emphasize their small, well-characterized measurement uncertainties in their homogeneous sample of 81 galaxies, and report an \rl\  scatter identical to that of C07. \citet{avila08} state that their heterogeneous sample of 76 galaxies spans a broad range of morphological types and surface brightnesses; they find  $\sigma_{logR} \sim 0.20 \log (h^{-1} \mathrm{kpc})$ for the \rl\ relation. \citet{shen03} study the \rl\ relation for a statistically complete sample of $\sim100\,000$ late-type galaxies from the SDSS, selected using cuts in light concentration and S\'ersic index. They report  $\sigma_{logR} \sim 0.13 \log (h^{-1} \mathrm{kpc})$ at the high-luminosity end of their relation (late and early types are likely mixed at the low-luminosity end; see \citealt{graham08}). We therefore find that relative to studies adopting a wide range of sample selection philosophies, the average scatter in the SFI++ \rl\ relation is {\it factors of 2.5 -- 4 smaller} than previously reported.

We attribute the significantly smaller scatter in our \rv\ and \rl\ relations to our use of homogeneously measured, extinction-corrected isophotal radii as disk sizes instead of the scale-lengths used in the above studies\footnote{ \citet{shen03} use Petrosian and S\'ersic half-light radii in their study, not $r_d$. Tests using the radius encompassing 83\% of the $I$-band light for SFI++ galaxies show that integral measures also produce scaling relations with more scatter than presented here  \citep{saintonge08}, presumably because of the inclusion of the bulge light and lack of extinction correction.}.  As explained in \S\ref{meas_rcor}, there are good reasons to expect that $R_{23.5}$ can be more reliably measured and corrected than $r_d$, and therefore that it should produce tighter scaling relations than the latter. We carry out a detailed analysis of the scatter in the \rl\ relation in \S\ref{budget}.

\begin{figure}
\epsscale{1.0}
\plotone{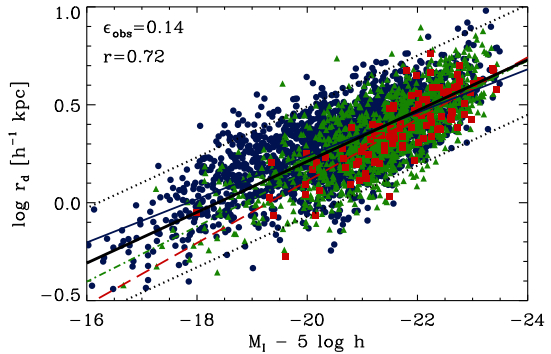}
\caption{The \rl\ relation for the nontemplate subsample, using the extinction-corrected disk scale-length ($r_d$) as the size indicator instead of $R_{23.5}$.  The data are broken down by morphological type, with lines and symbols described in Fig. \ref{morph_NT}.  The best orthogonal linear fit to the entire sample is given by the thick solid line, and the scatter $2\epsilon_{obs} $ about that fit is shown by the thick dotted lines. The values of $\epsilon_{obs}$ and  the Pearson correlation coefficient $r$ computed for the entire subsample are in the top-left corner. A color version of this figure is available in the electronic edition of the Journal.
\label{rd_NT}}
\end{figure}

  Fig.~\ref{rd_NT} shows the nontemplate \rl\ relation constructed in exactly the same manner as in Fig~\ref{VRL_NT}c, except that we use the SFI++ disk scale-length $r_d$, corrected for internal extinction using the prescription of \citet{giovanelli95}, instead of $R_{23.5}$. The values of $r$ and $\epsilon_{obs}$ of that relation (including all galaxy types) are given in the top-left corner of the plot. This version of the \rl\ relation has a similar $r$ to that found by C07, and $\epsilon_{obs}$ that is comparable to the scatters reported in the studies described above. Fig.~\ref{rd_NT} confirms: (1) the speculation by C07 that \rv\ and \rl\ relations constructed using $r_d$ are less robust than the \lv\ relation because $r_d$ is not reliably measured, (2) that extinction-corrected $r_{23.5}$ are superior to extinction-corrected $r_d$ when building scaling relations in that the former parameter yields significantly stronger correlations, and (3) that the small scatter in our \rl\ relations is not caused by selective ``pruning" of high-scatter points in the SFI++ \citep[][]{avila08}.

\begin{deluxetable*}{lccccccccc}
\tablecaption{Orthogonal Fit Parameters for SFI++ Scaling Relations\label{templates}}
\tablecolumns{10}
\tabletypesize{\footnotesize}
\tablehead{
\colhead{Relation} &  \multicolumn{4}{c}{Template}  &  \colhead{ } & \multicolumn{4}{c}{Non-template} \\
\cline{2-5} \cline{7-10}
& \colhead{$a \pm \delta a$}  & \colhead{$b \pm \delta b$} &
\colhead{$\sigma$} &  \colhead{$r$} &\colhead{ } & \colhead{$a \pm \delta a$}  & \colhead{$b \pm \delta b$} & \colhead{$\sigma$} & \colhead{$r$}\\
\hline \\
\multicolumn{10}{c}{NO MORPHOLOGICAL CORRECTION}}
\startdata
\textit{\textbf{LV}}\tablenotemark{a}:  &
$-20.83\pm0.17$  & $
-8.79 \pm 0.04$ &
0.55 & 0.91&  &
$-20.77\pm0.07$  & $
-9.07 \pm 0.02$ &
0.54 & 0.87\\
\textit{\textbf{RV}}\tablenotemark{b}:  &
$0.911\pm0.037$  & $
1.281 \pm 0.011$ &
0.11 & 0.84&  &
$0.910\pm0.015$  & $
1.357 \pm 0.004$ &
0.11 & 0.83\\
\textit{\textbf{RL}}\tablenotemark{c}:  &
$0.763\pm0.028$  & $
-0.159 \pm 0.001$ &
0.05 & 0.96&  &
$0.763\pm0.016$  & $
-0.165 \pm 0.001$ &
0.05 & 0.97\\

\cutinhead{WITH MORPHOLOGICAL CORRECTION}

\textit{\textbf{LV}}\tablenotemark{a}:  &
$-20.88\pm0.15$  & $
-9.24 \pm 0.04$ &
0.55 & 0.91&  &
$-20.81\pm0.06$  & $
-9.42 \pm 0.01$ &
0.53 & 0.88\\
\textit{\textbf{RV}}\tablenotemark{b}:  &
$0.927\pm0.038$  & $
1.353 \pm 0.011$ &
0.11 & 0.85&  &
$0.922\pm0.019$  & $
1.407 \pm 0.005$ &
0.11 & 0.84\\
\textit{\textbf{RL}}\tablenotemark{c}:  &
$0.774\pm0.029$  & $
-0.160 \pm 0.001$ &
0.05 & 0.97&  &
$0.773\pm0.012$  & $
-0.164 \pm 0.000$ &
0.05 & 0.97
\enddata
\tablenotetext{a}{The \lv\ relation is parametrized as $M_I - 5\log h = a + b(\log W -2.5)$. }
\tablenotetext{b}{The \rv\ relation is parametrized as $\log R \, [h^{-1}\,\mathrm{kpc}] = a + b(\log W -2.5)$. }
\tablenotetext{c}{The \rl\ relation is parametrized as  $\log R \, [h^{-1}\,\mathrm{kpc}]  = a + b(M_{I}+20)$.}
\end{deluxetable*}

\subsection{Morphological Dependence \label{morphology}}

\begin{figure}
\epsscale{0.9}
\plotone{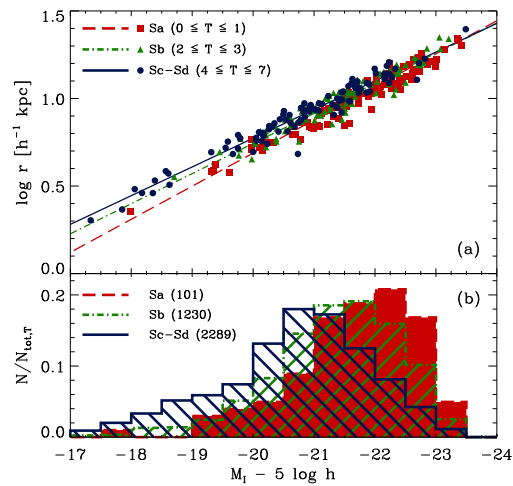}
\caption{{\it a)} \rl\ relation for the nontemplate subsample, subdivided by morphological type.  The lines represent the best orthogonal linear fits to the different subsets. To avoid overcrowding, only 100 randomly drawn galaxies of each morphological type are plotted. The definition of each morphological class based on the T-types of S07 is given in the upper-left corner.  {\it b)} Normalized histograms showing the luminosity distribution of each morphological type, with the total number of galaxies of each type given. A color version of this figure is available in the electronic edition of the Journal.
\label{morph_NT}}
\end{figure}

As discussed by M06, the SFI++ contains galaxies with a broader range of morphological types than earlier incarnations of the catalog, affording a more thorough investigation of scaling relations as a function morphological type. Nonetheless, the histograms in  Fig.~\ref{morph_NT}b illustrate that the SFI++ is dominated by Sc galaxies.  As such, it is difficult to disentangle sample biases from physical effects in any scaling relation differences that are found.  We therefore follow the approach of \citet{giovanelli97} and M06, and adjust the best-fitting scaling relations determined for galaxy types Sa and Sb to match that of the best-fitting Sc+Sd relations. 
We use the larger nontemplate subsample for this exercise, and apply the derived correction to both the template and nontemplate subsamples. 

The computed slopes, zero-points and scatters of the scaling relations to which these morphological corrections have and have not been applied are given in Table~\ref{templates}.  There is little difference between the properties of the uncorrected and corrected relations: this stems from the high Sc fraction in the SFI++. For this reason, none of the conclusions in our study are affected by our application of the correction. 
The morphological dependence of the \lv\ relation has been extensively discussed in the literature \citep[e.g.][C07]{roberts78,aaronson83,rubin85,giraud86,pierce88,kannappan02}. That of the SFI++ \lv\ relation in particular has already been addressed by \citet{giovanelli97} and M06. We therefore focus on the morphological dependence of the \rl\ relation, and provide corrections for the \lv\ and \rv\ relations using the same method. 
 
 Fig.~\ref{morph_NT}a shows the \rl\ relations for nontemplate galaxies of different morphological types, as classified in Tables~2~and~4 of S07.  Note that for clarity, we only plot a random subset of 100 galaxies of each morphological class. The lines in Fig.~\ref{morph_NT}a show the best orthogonal linear fits to galaxies of each morphological type. We detect a clear trend: at a given $L$, early-type spirals have smaller $R$ than late-type spirals, and the slope of the early-type \rl\ relation is steeper. This trend is qualitatively similar to that found by \citet{shen03} for their complete sample.
 Based on the orthogonal fits in Fig.~\ref{morph_NT}a, we apply the following additive factor to $L$ in the \rl\ relation so that the mean relation for early types matches that of the later types:
 \begin{equation}
\Delta _{RL}= \Bigg \{ \begin{array}{ll}
 0.57+0.024(M_I+20)  & \mbox{for types S0a/Sa}\\
 0.18+0.008(M_I+20) & \mbox{for types Sab/Sb}\\
 0.0 & \mbox{for later types}.
 \end{array}
 \label{deltaRL}
\end{equation}
Note that this correction is only applied to $L$ in the \rl\ relation; $L$ for individual Sa and Sb galaxies in the \lv\ and \rl\ relations of Figs.~\ref{VRL_T}~and~\ref{VRL_NT} are thus slightly different by virtue of the morphological corrections applied.  We include a 15\% uncertainty in this correction in the \rl\ relation error budget (see Appendix~\ref{appendix2}).  

 We find a lesser dependence of the \rl\ relation parameters on morphology than reported by C07. This difference stems from our definition of $R$: the lines in Fig.~\ref{rd_NT} show the best-fitting relations constructed with $r_d$ for different morphological types, where we recover a strong dependence. The systematic variation of $\langle r_d/r_{23.5} \rangle$ in Fig.~\ref{rd_r235} can explain at least part of the difference in morphological dependence between these two indicators: relative to $r_d$, $r_{23.5}$ is systematically larger for brighter galaxies, which are more likely to be Sa's according to Fig.~\ref{morph_NT}b. This works to offset the strong trend found when $r_d$ is used to construct the \rl\ relation. However, we cannot rule out systematic effects in how $r_d$ is measured in Sa's relative to Sc's, such as a bias in the manner that Sa disks are marked given their more substantial bulges (see \S\ref{meas_rcor}).
 
 We derive the morphological corrections to the \rv\ and \lv\ relations in an analogous manner to that for the \rl\ relation:
  \begin{equation}
\Delta _{RV}= \Bigg \{ \begin{array}{ll}
 0.77-0.25(\log W - 2.5)  & \mbox{for types S0a/Sa}\\
 -0.13+0.06(\log W - 2.5) & \mbox{for types Sab/Sb}\\
 0.0 & \mbox{for later types}.
 \end{array}
 \label{deltaRV}
\end{equation}
\begin{equation}
\Delta _{LV}= \Bigg \{ \begin{array}{ll}
 -1.24+0.31(\log W - 2.5)  & \mbox{for types S0a/Sa}\\
 2.54-1.01(\log W - 2.5) & \mbox{for types Sab/Sb}\\
 0.0 & \mbox{for later types}.
 \label{deltaLV}
 \end{array}
\end{equation}
The small difference between the correction in eq.~\ref{deltaLV} and that from M06 is likely due to the fitting techniques (bisector fits in M06 and orthogonal fits with $x$ and $y$ errors here) and to sample selection: we calculate our corrections using the larger nontemplate subsample, while M06 use their incompleteness-corrected SFI++ template relation.

\subsection{Scaling Relation Residuals \label{residuals}}

The residuals of the scaling relations provide useful checks on data quality as well as important insight into the nature of the the relations themselves. Here, we exploit the large size of the nontemplate subsample to examine the residuals of the scaling relations presented in \S\ref{vrl}.

\begin{figure}
\epsscale{1.0}
\plotone{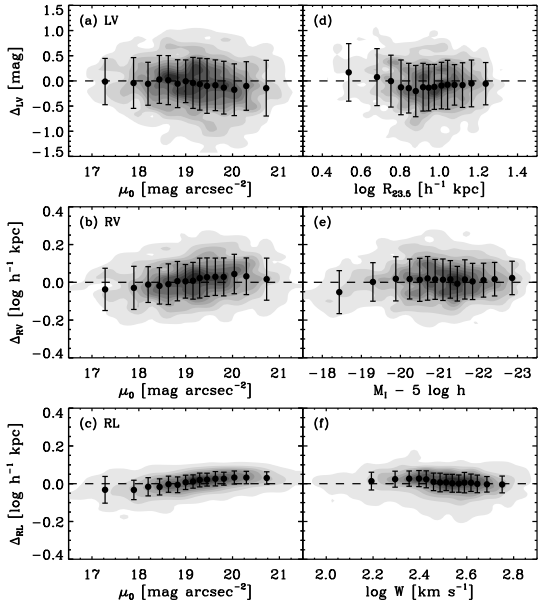}
\caption{Data - model residuals of the scaling relations for the nontemplate subsample, plotted as a function of central $I$-band surface brightness {\it (a, b, c)} and  the third scaling parameter {\it (d, e, f)}, namely $R$ for the \lv\ relation, $L$ for the \rv\ relation and $V$ for the \rl\ relation.  In all panels, contours show the distribution of all datatpoints. The red points show the median 
value in equally populated bins, with errorbars representing the 1$\sigma$ spread in the distribution of values. \label{res_fig}}
\end{figure}

Fig.~\ref{res_fig} plots the data-model residuals of the \lv, \rv\ and \rl\ relations as a function of $I$-band central surface brightness $\mu_0$ \citep[derived as in][]{haynes99} and the third scaling relation parameter. In all panels, the residual is computed in the variable on the $y$-axis in Fig.~\ref{VRL_NT}. We find no significant correlations between any of the scaling relation residuals and these parameters. We have also checked for correlations with variables such as inclination that might imply biases in our samples, and find none.

 The lack of correlation between the residuals of the \lv\ relation and $\mu_0$ has been extensively discussed in the literature \citep{sprayberry95,zwaan95,courteau99,firmani00,vdb00,kannappan02}. It is generally interpreted as evidence in favor of sub-maximal disks \citep[e.g.][]{bershady10}, but \citet{dutton07} demonstrate that a variety of factors can alter the surface brightness dependence of the \lv\ relation \citep[see also][]{firmani00,vdb00}. 
  
\begin{figure}
\epsscale{1.0}
\plotone{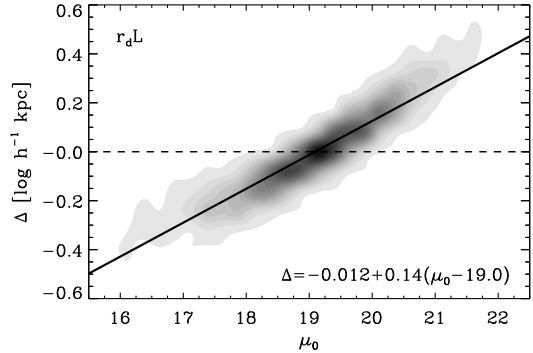}
\caption{Residuals of the \rl\ relation for the nontemplate sample constructed using the extinction corrected disk scale-length instead of $R_{23.5}$ (Fig.~\ref{rd_NT}), plotted against the $I$-band central surface brightness. The best-fitting linear relation is shown by the solid line, and is given in the bottom-right corner. 
\label{res_rd_fig}}
\end{figure}

 On the other hand, the lack of correlation between the residuals of the \rl\  and \rv\ relations and $\mu_0$ is not typical of past survey results. The reason is again our definition of $R$: Fig.~\ref{res_rd_fig} shows that we recover a strong surface brightness correlation when the extinction-corrected scale-length is used as the measure of disk size in the \rl\ relation, with a slope consistent with that reported by C07. In analogy to the morphological dependence discussed in \S\ref{morphology}, the change in $\langle r_d/r_{23.5} \rangle$ as a function of $L$ in Fig.~\ref{rd_r235} can also explain the lack of correlation when $R_{23.5}$ is used to construct the \rl\ relation: brighter, higher surface brightness disks have lower $\langle r_d/r_{23.5} \rangle$, which reconciles Fig.~\ref{res_rd_fig} and Fig.~\ref{res_fig}c.  However, it is also possible that systematic effects related to the measurement of $r_d$ and $\mu_0$ (the latter an extrapolation of the exponential disk defined by the former) also influence the trend in Fig.~\ref{res_rd_fig}. 

\begin{figure*}
\epsscale{1.0}
\plotone{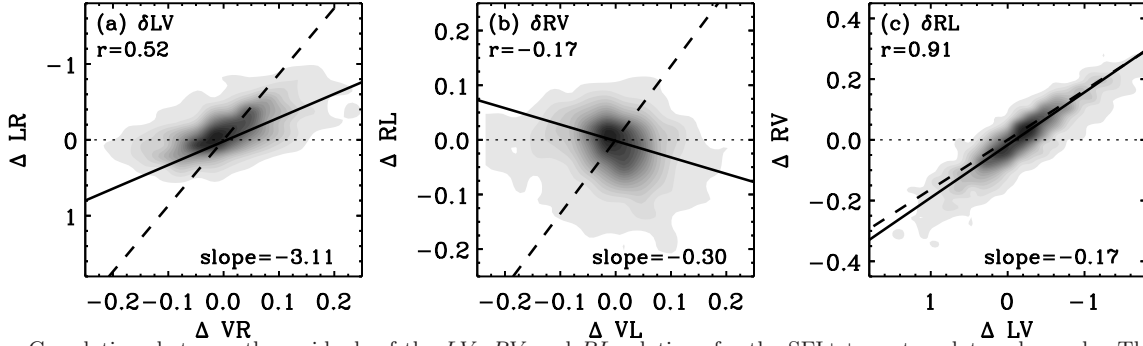}
\caption{Correlations between the residuals of the \lv, \rv\ and \rl\ relations for the SFI++ nontemplate subsample. The axis labels denote how the residual is defined in each case: for example, $\Delta LR$ in $(a)$ are data - model residuals in $L$ computed from the \rl\ relation, whereas $\Delta RL$ in $(b)$ are data - model residuals in $R$ computed from the \rl\ relation. Panels $(a)$, $(b)$ and $(c)$ therefore represent the \lv, \rv\ and \rl\ relations in differential form, respectively.  
In each panel, the Pearson correlation coefficient of the residuals is in the top-left corner, the solid lines are the best-fitting linear relations to the residuals, and their slope is in the bottom-right corner. The dashed lines represent the slope of the corresponding scaling relation from Fig.~\ref{VRL_NT}. 
\label{cor_res}}
\end{figure*}

Finally, we investigate the correlation between the scaling relation residuals, shown in Fig.~\ref{cor_res}. The axis labels in Fig.~\ref{cor_res} reflect the definition of the residuals: for example, $\Delta LR$ on the $y$-axis of Fig.~\ref{cor_res}a are the residuals of the \rl\ relation, computed as data - model residuals in $L$, whereas $\Delta RL$ on the $y$-axis of Fig.~\ref{cor_res}b are the residuals of the \rl\ relation,  computed as data - model residuals in $R$.  Figs~\ref{cor_res}a,~\ref{cor_res}b~and~\ref{cor_res}c therefore correspond to the \lv, \rv\ and \rl\  relation in differential form. 

As in P05 and C07, we find little evidence for a correlation between the \lv\ and \rl\ relation residuals in Fig.~\ref{cor_res}b. 
\citet{courteau99} argue that this lack of correlation suggests that even high surface brightness galaxies have submaximal disks. However, the models of \citet{gnedin07} and \citet{dutton07} demonstrate that factors such as bulge formation, stellar and gas fractions, scatter in halo properties and disk mass-to-light ratios also influence the behavior of the residuals. 
We find a weak positive correlation between the \rl\ and \rv\ relation residuals, and a strong positive correlation between the \lv\ and \rv\ relation residuals, whose slopes are in rough agreement with those of the actual \lv\ and \rv\ relations, respectively.  As explained in C07, these results are qualitatively consistent with little scatter in $L$ and uncorrelated scatter in $V$ and $R$. The error budget analysis in \S\ref{budget} supports these conclusions for the measurement errors in $V$, $L$ and $R$, but the substantial intrinsic scatter in all of the relations requires detailed modeling in order to fully characterize the origin of the residual correlations.

\section{Error Budgets of the Scaling Relations}
\label{budget}

An understanding of the observed scaling relation scatter $\epsilon_{obs}$ is essential for reconstructing large-scale structure from extracted peculiar velocities \citep[e.g.][]{sandage75,giraud86vf,dale99b} as well as for modeling the origin of the relations themselves \citep[e.g.][see \S\ref{intro}]{dutton07,gnedin07}. In the same spirit as the analyses carried out by \citet{giovanelli97} and M06 for the SFI++ template \lv\ relation, we compute the error budgets for both the template and nontemplate \lv, \rv\ and \rl\ relations presented in \S\ref{scalingrelations}. 

The total observed scatter in the scaling relations, $\epsilon_{obs}$, is a combination of the measurement uncertainties $\epsilon_{mes}$ and $\epsilon_{int}$, the intrinsic scatter:
\begin{equation}
\epsilon_{obs}^2=\epsilon_{mes}^2+\epsilon_{int}^2
\label{scat_obs}
\end{equation}

For each relation,  $\epsilon_{mes}$ is computed from the measurement uncertainty on the two scaling parameters and their covariance. We also include a contribution from the morphological corrections of \S\ref{morphology}.  
The mathematical details of the derivation of $\epsilon_{mes}$ for the three scaling relations can be found in Appendix \ref{appendix2}.

\subsection{The \lv\ and \rv\ relations}
\label{budget_lv}

\begin{figure}
\epsscale{0.9}
\plotone{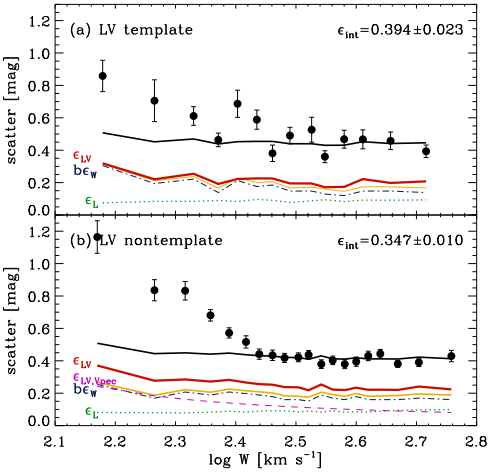}
\caption{Error budget of the \lv\ relation for the SFI++ template $(a)$ and nontemplate $(b)$ subsamples. Data points are the median observed scatter ($\epsilon_{obs}$) as a function of $V$, and the error bars are the uncertainty on the position of the median.  The separate contributions to the total scatter are:  the measurement error on $L$ ($\epsilon_{L}$, dotted line), the measurement error on $V$ ($\left| b \right| \epsilon_{W}$, dash-dotted line), which add in quadrature to produce the light solid line (i.e. neglecting the contribution of correlated errors).   The total measurement error $\epsilon_{mes} \equiv \epsilon_{LV}$ (eq. \ref{epsilonLV}, lower dark solid line) accounts for correlated measurement uncertainties and, for the nontemplate subsample $(b)$, the contribution of a peculiar velocity of amplitude $V_{pec}=300\,\kms$ ($\epsilon_{LV,Vpec}$, dashed line).  The solid line that goes through the points at $\log W > 2.4$ is the sum in quadrature of $\epsilon_{LV}$ and of a constant intrinsic scatter $\epsilon_{int}$. A color version of this figure is available in the electronic version of the Journal. 
\label{budget_LV}}
\end{figure}

\begin{figure}
\epsscale{0.9}
\plotone{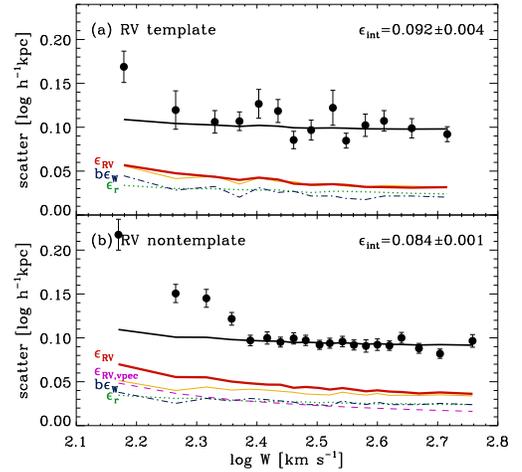}
\caption{Error budget of the \rv\ relation for the SFI++ template $(a)$ and nontemplate $(b)$ subsamples.  Lines and symbols are as in Fig. \ref{budget_LV}. A color version of this figure is available in the electronic version of the Journal. 
\label{budget_RV}}
\end{figure}

Figures \ref{budget_LV} and \ref{budget_RV} show the total scatter of the \lv\ and \rv\ relations as a function of $V$ for both SFI++ subsamples.  We note that our estimates of $\epsilon_{mes}$ (dash-dotted and dotted lines) for each variable are similar to that reported in other studies \citep[e.g.][]{dutton07,gnedin07}, and that accounting for correlated uncertainties does not significantly change $\epsilon_{mes}$ \citep[lower dark and light solid lines; see also][]{giovanelli97}.  

In all cases, the amount of scatter is almost constant at $\log W \gtrsim 2.4$, but increases at lower $V$. Above this threshold, $\epsilon_{obs}$ can be very well reproduced by adding a constant intrinsic scatter $\epsilon_{int}$ in quadrature with $\epsilon_{mes}$ ($\epsilon_{LV}$ and $\epsilon_{RV}$; the red line in each panel).  

The value of $\epsilon_{int}$ required to match $\epsilon_{obs}$ at $\log W > 2.4$ for the template and nontemplate subsamples are broadly consistent, and we attribute the slightly larger value in the template case to a small peculiar velocity effect in that subsample (see \ref{pec_vel}).  Using the values derived from the nontemplate subsample, which benefits from much better statistics, we adopt intrinsic scatters $\epsilon_{LV,int}=0.35\pm0.01$ mag and $\epsilon_{RV,int}=0.084\pm0.001 \log ($h$^{-1}$ kpc$)$ for the \lv\ and \rv\ relations, respectively. 

Our value of $\epsilon_{LV,int}$ is similar to that invoked by M06 in their error budget of the SFI++ template \lv\ relation, although they fit for a width-dependent scatter including points with $\log W < 2.4$.  That solution is perfectly consistent with our error budget of Fig.~\ref{budget_LV}a, but the cleaner view provided by the large nontemplate subsample (Fig.~\ref{budget_LV}b) makes us favor the constant scatter approach.

 As for the deviation from constant intrinsic scatter at low $V$ in Figs.~\ref{budget_LV}~and~\ref{budget_RV}, we identify four possible causes: peculiar velocities, sample incompleteness, increased measurement errors and increased intrinsic scatter.  We investigate these possibilities in turn below.

\subsubsection{Peculiar velocities}
\label{pec_vel}
Peculiar velocities introduce scatter in both the \lv\ and \rv\ relations.   Because distances to nontemplate galaxies are derived assuming pure Hubble flow, the effect will be strongest in that subsample.  While peculiar velocities in clusters are larger than in the field, the template sample is mostly free of peculiar velocity-induced scatter, because we adopt cluster distances for all their members (see \S \ref{distances}).

The median redshift $\overline{cz}$ of the galaxies in the nontemplate subsample is a strong function of $V$.  Since a galaxy at redshift $cz$ with peculiar velocity $V_{pec}$ will scatter away from the mean \lv\ and \rv\ relations in proportion to $\log (1+V_{pec}/cz)$ (see Appendix~\ref{appendix2}), the effect of peculiar velocities on the scatter in these relations is strongest at low $V$, where $\overline{cz}$ is smallest. 

 To assess the impact of peculiar velocities, we compute the scatter produced in the \lv\ and \rv\ relations by a characteristic $V_{pec} = 300\,\kms$ and $\overline{cz}$ as a function of $V$ in the nontemplate subsample. The result is illustrated by the dashed lines in Figs.~\ref{budget_LV}b~and~\ref{budget_RV}b. At the low-$V$ end of the nontemplate relations, peculiar velocities can constitute the largest source of scatter. The lower dark solid line in these figures adds this contribution to the scatter in quadrature with $\epsilon_{mes}$. The resulting is a steepening of the total measurement error curve at low $V$.  However, it is clear that peculiar velocities alone cannot explain the increase in $\epsilon_{obs}$ in the \lv\ and \rv\ relations at low $V$.

\subsubsection{Sample incompleteness} 
\label{incompleteness}

In addition to biasing scaling relation slopes and zero points (see \S\ref{vrl}), statistical incompleteness of a sample can also affect their scatter.  Cluster samples are often used for scaling relation studies because they are less affected by incompleteness than flux-limited field samples.  Nonetheless, \citet{giovanelli97} and M06 have shown that the SFI++ template subsample suffers from incompleteness at $\log W \lesssim 2.5$ (e.g.\ Fig.\ 4 in M06).  

The nontemplate subsample certainly also suffers from incompleteness. However, the union of several disparate samples to create the SFI++ makes it impossible to derive the selection function of the survey and quantify its incompleteness (see \S\ref{intro} and the discussion in S07).  It is nonetheless reasonable to assume that the qualitative impact of incompleteness mirrors that determined by M06 \citep[see also][]{dejong00}.  Since we are more likely to observe larger/brighter galaxies near the survey limits, the scatter at the low-$V$ end of the relations is reduced. However, we observe an {\it increase} in $\epsilon_{obs}$. Thus unless incompleteness in our SFI++ subsamples behaves in the opposite manner from that determined by M06 for the SFI++ template \lv\ relation, it is unlikely to explain the increase in $\epsilon_{obs}$ at $\log W \lesssim 2.4$ in Figs.~\ref{budget_LV}~and~\ref{budget_RV}.

\subsubsection{Increased scatter in $V$}

Contrary to the \lv\ and \rv\ relations, $\epsilon_{obs}$ in the \rl\ relation does not increase at low $L$ (see \S \ref{RL_budget} and Fig. \ref{budget_RL}).  For this reason, we associate the behavior of the scatter at low $V$ in the \lv\ and \rv\ relations to $V$ itself. Here, we investigate the possibility that there is an additional contribution to the scatter in $V$ for which we have not accounted in our error budget. 

Part of the increased scatter could come from measurement uncertainties specific to galaxies with low $V$.
Galaxies with narrower HI lines are more prone to ``catastrophic" measurement errors due to, for example, line asymmetries and turbulence effects \citep{springob05}, or uncertainties in inclination.  
While it is impossible to account quantitatively for the effect of these catastrophic errors on our scaling relations, it is plausible that they produce a total measurement error function ($\epsilon_{LV}$, $\epsilon_{RV}$) that is steep at the low-$V$ end.
 Nonetheless, it seems unlikely that the median fractional uncertainty on $\log W$ at $\log W \sim 2.2$ is underestimated by a factor of 4 relative to that computed in our budget, and thus that measurement errors alone explain the increase in $\epsilon_{obs}$.

We are therefore left with the possibility that the scatter at low $V$ stems from an astrophysical effect. For example, galaxies with low $V$ have lower masses, and are therefore more susceptible to HI disk stripping processes.  Since these galaxies tend to have rising rotation curves \citep[e.g.][]{catinella06}, measuring $V$ at different radii because of HI disks stripped to various degrees will introduce scatter.    
However, the slope of low-mass galaxy rotation curves is not nearly steep enough to explain the entirety of the increase in $\epsilon_{obs}$. 

It is possible that the behavior of $\epsilon_{obs}$ in the \lv\ and \rv\ relations stems from an increase in the intrinsic scatter at low $V$ related to the properties of their stellar disks or their parent dark matter halos. While this is a tantalizing possibility at the outset, it is difficult to ``tune" galaxy formation models of the type described in \S\ref{gal_form} to increase the scatter in the \lv\ and \rv\ relations without increasing that in the \rl\ relation \citep[e.g.][]{dutton07}.  It is therefore not obvious that the increase in $\epsilon_{obs}$ at $\log W < 2.4$ in the \lv\ and \rv\ relations can be attributed to intrinsic scatter, although a full suite of galaxy formation models is needed to address this issue.

\subsection{The \rl\ relation}
\label{RL_budget}

\begin{figure}
\epsscale{0.9}
\plotone{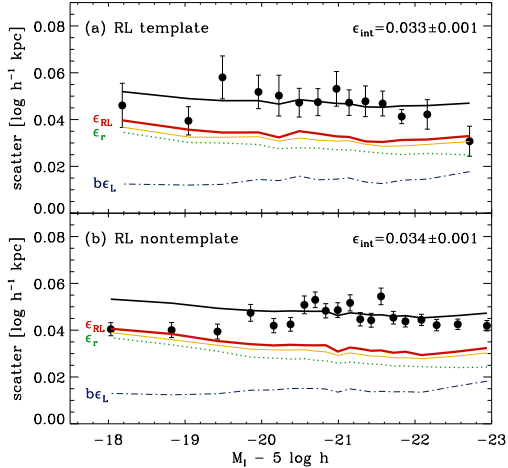}
\caption{Error budget of the \rl\ relation for the SFI++ template $(a)$ and nontemplate $(b)$ subsamples. Lines and symbols are as in Fig. \ref{budget_LV}. In this case we do not include the contribution of peculiar velocities to the total scatter, because as discussed in Appendix \ref{appendix2}, they move points almost exactly along the \rl\ relation. A color version of this figure is available in the electronic version of the Journal.
\label{budget_RL}}
\end{figure}

The average $\epsilon_{obs}$ for the \rl\ relation is significantly lower than that of the \rv\ relation (Table~\ref{templates}).  This comes in part because the measurement errors on $L$ introduce less scatter than those of $V$, as pointed out by C07 and clearly seen in the error budgets of Figures \ref{budget_RV} and \ref{budget_RL}.  However, our analysis shows that the relation also has significantly less intrinsic scatter: in regions at high $V$ and high $L$ where the scatter is approximately constant, we measure $\epsilon_{RV,int}=0.084\pm0.001 \log(h^{-1}\mathrm{kpc})$ for the \rv\ relation but $\epsilon_{RL,int}=0.034\pm0.001 \log(h^{-1}\mathrm{kpc})$ for the \rl\ relation.  

The overall behaviour of the scatter in the \rl\ relation is also different: contrary to the increases found for the \lv\ and \rv\ relations, we find a {\it reduced} scatter for $M_I - 5 \log h \gtrsim -20$ in the \rl\ relation.  We can rule out peculiar velocities affecting the scatter since they move points almost exactly along the \rl\ relation (see Appendix \ref{appendix2}).   Given the discussion in \S\ref{incompleteness}, incompleteness is the most likely culprit.  Since some of the data included in the SFI++ stem from aperture diameter-limited surveys (especially among the fainter, more nearby galaxies), we expect incompleteness to reduce the scatter at the faint end of the \rl\ relation in analogy the \lv\ relation behaviour determined by M06 for the SFI++ template.     

\section{Applications of the \rl\ and \rv\ relations}
\label{applications}

In \S\ref{vrl} we present \rv\ and \rl\ relations that exhibit significantly smaller scatter than has been previously reported, which  we attribute to our adoption of an inclination-corrected isophotal radius as the size parameter $R$ in these relations rather than a disk scale-length. Here, we discuss the implications of this reduced scatter for two common scaling relation applications: redshift-independent distances and galaxy formation models. 
 
 While it is obvious that the ideal samples to be used as distance indicators and galaxy formation constraints differ, it is equally obvious that these ideal samples are not yet available. The use of the scaling relations as either distance indicators or galaxy formation constraints thus requires finding an acceptable middle ground with respect to sample selection.  It is in this sense that subsets of the SFI++ are relevant in both the distance indicator and galaxy formation contexts. The SFI++ template sample was explicitly constructed to measure \lv\ distances (M06), and we discuss in \S\ref{dist_ind} the potential of exploiting our template \rv\ relation for the same purpose. The SFI++ nontemplate galaxies span a broader range of properties, and were not selectively edited for cosmic flows or other purposes. The mix of morphological types in the SFI++, as with most other large catalogs of similar quality, is not representative of that in the local Universe. On the other hand, there is no reason to suspect that the SFI++ does not reflect the properties of Sc galaxies in that volume. With this caveat in mind, we explore  in \S\ref{gal_form} the implications of the small scatter in the nontemplate \rl\ relation for galaxy formation models.

 Note that in this section, we adopt the {\it WMAP5} value of $H_0=71.9^{+2.6}_{-2.7}$ km s$^{-1}$ Mpc$^{-1}$ \citep{dunkley09} for concreteness.

\subsection{The \rv\ Relation as a  Distance Indicator \label{dist_ind}}

Since $R$ depends on distance while $V$ does not, the \rv\ relation could in principle be used to produce distance-independent redshifts just like the widely-used \lv\ relation. This application has received little attention so far, because the \rv\ relation typically has significant scatter.  
However, our \rv\ relation is nearly as tight as the \lv\ relation, and has $\sim1.5$ times less scatter than found in previous studies (see Figs.\ \ref{VRL_T} and \S \ref{scalingrelations}). Since disk sizes can be more straightforward to compute than magnitudes (and sometimes even more reliable), there is obvious interest in using the \rv\ relation as a distance indicator.

As a proof of principle, we use the sample of 17 SFI++ galaxies with a distance measurement from Cepheid variables compiled by M06 to evaluate the distances derived from the SFI++ \rv\ relation. We use the template subsample \rv\ relation for this exercise, since it is less susceptible to peculiar velocities and  can in principle be corrected for incompleteness. The template sample in M06 was used to independently calibrate the zero point of the \lv\ relation and to provide an estimate of $H_{0}$. Their value of $H_{0}=74\pm2 \,\mathrm{(random)} \pm 6 \,\mathrm{(systematic)}$ km s$^{-1}$ Mpc$^{-1}$ compares well with the {\it WMAP5} result of $H_0=71.9^{+2.6}_{-2.7}$ km s$^{-1}$ Mpc$^{-1}$ \citep{dunkley09}. The systematic component of the uncertainty in the M06 estimate stems largely from the uncertainty of the Cepheid distance zero point, not the \lv\ relation \citep{sakai04}.

\begin{figure}
\epsscale{1.2}
\plotone{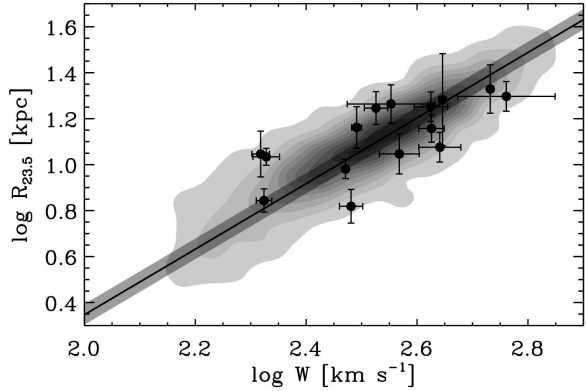}
\caption{\rv\ relation for galaxies with distance measurements from Cepheid variables (points).  The filled contours shows the template \rv\ relation normalized to $H_0=71.9$ km s$^{-1}$ Mpc$^{-1}$, and the solid line is the best fit to these data. The shaded region shows the 1$\sigma$ uncertainty in the zero point of the Cepheid \rv\ relation when its slope is held fixed to the template \rv\ relation value.  
\label{RVceph}}
\end{figure}

Fig.~\ref{RVceph} shows the \rv\ relation for the Cepheid sample, overlaid on the template \rv\ relation normalized to the {\it WMAP5} value of $H_0$.   Since most of the galaxies in the Cepheid sample have large rotation velocities, it is not possible to reliably fit the slope of their \rv\ relation. Instead, we adopt the slope of the template \rv\ relation, but determine the zero point of the Cepheid sample \rv\ relation independently.  The 1$\sigma$ region around the \rv\ relation with this Cepheid-derived zero point is shown in Fig.~\ref{RVceph} as the shaded region, which overlaps comfortably with that derived from the template subsample (solid line).  

Reversing the problem, the value of $H_0$ can be determined by matching the zero points of the SFI++ template relation and that of the cosmology-independent value derived from the Cepheid sample. We find a value of $H_0=72\pm7$ km s$^{-1}$ Mpc$^{-1}$, where the errors come in part from the intrinsic scatter of the \rv\ relation but mostly from the uncertainty on the Cepheid distances. Comparing the M06 estimate of $H_0$ with this value, we conclude that the \lv\ and \rv\ relations provide similar constraints on this parameter.

\begin{figure}
\epsscale{1.1}
\plotone{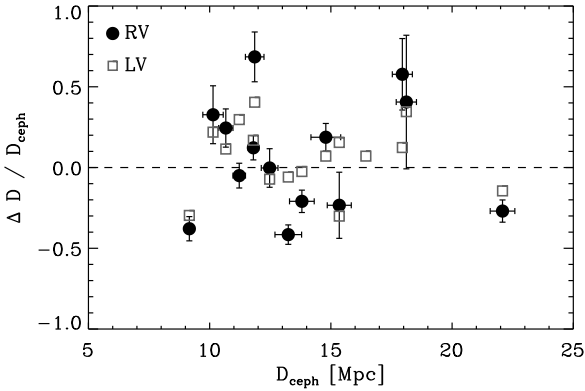}
\caption{Relative difference between the distances from the Cepheid variable measurements and the distances derived from the SFI++ scaling relations (the \rv\ and \lv\ relations, represented as filled circles and open squares, respectively). 
\label{dist_ceph}}
\end{figure}

Finally, the distances to the galaxies in the Cepheid sample are computed from the template subsample \lv\  and \rv\ relations in \S \ref{vrl}.  The relative differences between the Cepheid- and scaling relation-derived distances are shown in Figure \ref{dist_ceph}.   The typical errors on the \lv - and \rv -derived distances are obtained by taking the median absolute deviation of these relative differences, and are 14.8\% and 24.7\%, respectively.   These errors are in line with the typical errors of $\sim 15-20$\% generally quoted for \lv\ distances \citep[e.g.][]{tully00}.

The above exercises suggest that even given the tight \rv\ relation presented here, the \lv\ relation produces more accurate redshift-independent distances. However, in situations where accurate luminosities are not available, a well-defined \rv\ relation can be substituted with only moderately larger uncertainties. 
We note that a thorough analysis of the \rv\ relation as a distance indicator requires a correction for incompleteness effects. This is clearly feasible for the SFI++ template sample \citep[][M06]{giovanelli97}, but we defer this task to a future paper.

\subsection{The \rl\ Relation and Galaxy Formation Models}
\label{gal_form}

It is clear that a simultaneous application of galaxy formation models to the scaling relations derived here, like that carried out by \citet{dutton07} and \citet{gnedin07}, is beyond the scope of this paper. Nonetheless, a consideration of the SFI++ \rl\ relation scatter in the context of a simple model demonstrates the potential of our relations to place new constraints on galaxy formation. The \rl\ relation is ideal in this context because peculiar velocities move points almost exactly along the relation, and therefore do not contribute to the scatter (see Appendix~\ref{appendix2}). We use the larger nontemplate subsample for this exercise, with the morphological correction derived in \S\ref{morphology} applied. This relation therefore constrains the formation and evolution of present-day Sc galaxies.

\subsubsection{Model Details \label{galmod_model}}
We consider a \citet[][hereafter MMW98]{mo98} - style model of the SFI++ \rl\ relation and its scatter, where a self-gravitating exponential disk is embedded in an NFW halo.  Specifically, we eliminate the halo velocity $V_{200}$ from their eqs.~16 and 28 to produce the predicted relationship between the disk luminosity $L_{d}$ and scale-length $R_d$ of the stellar mass distribution:
\begin{equation}
R_d = \left[ \frac{G}{200\sqrt{2}H_0^2} \right]^{1/3} \, \frac{\Upsilon_d^{1/3} \lambda^\prime}{m_d^{1/3}}\, L_d^{1/3} f_c^{-1/2} f_R\,\,\,.
\label{galmodr}
\end{equation}

In eq.~\ref{galmodr}, $\Upsilon_d$ is the disk mass-to-light ratio. 
We adopt the relation derived by \citet{dutton07} using the population synthesis models of \citet{bell03} and a ``diet Salpeter" IMF:
\begin{equation}
\log \left(\frac{\Upsilon_{d,I}}{\rm{M_\odot/L_{\odot,I}}} \right) = 0.172 + 0.144 \log \left( \frac{L_{d,I}}{10^{10.3}\,\rm{L_{\odot,I}}} \right) \,\,\,.
\label{ml}
\end{equation}

The parameter $m_d$ in eq.~\ref{galmodr} is the ratio of the disk mass $M_d$ to the halo mass $M_{200}$.  Few constraints exist on $m_d$ or on its scatter at fixed $M_{200}$, but most spiral galaxy mass models find $m_d \sim 0.05$. As in \citet{dutton07}, we allow $m_d$ to vary with disk mass in order to fit the slope of the \rl\ relation:
\begin{equation}
m_d = m_{d,0} \left( \frac{L_{d,I}}{10^{10.3} \,\rm{L_{\odot,I}}} \right)^\alpha\,\,\,,
\end{equation}
where $m_{d,0}$ and $\alpha$ are free parameters. We tune $m_{d,0}$ and $\alpha$ to match the \rl\ relation intercept and slope, respectively.

 In eq.~\ref{galmodr}, $\lambda^\prime$ is the effective spin parameter of the system:
\begin{equation}
\lambda^\prime = \frac{j_d}{m_d} \lambda = \left( \frac{j_d}{m_d} \right) \frac{J_{200} |E|^{1/2}}{GM_{200}^{5/2}}  \,\,\,,
\end{equation}
where $j_d$ is the ratio of the disk angular momentum $J_d$ to the halo angular momentum $J_{200}$, and $\lambda$ is the spin parameter of the halo. Collisionless simulations of halo assembly find a log-normal distribution of $\lambda$ that is independent of halo mass. We adopt the distribution peak $\bar{\lambda} = 0.042$ \citep{bullock01} as our fiducial value for $\lambda^\prime$. The relaxed halos in the {\it WMAP5} simulations of \citet{maccio08} have a logarithmic scatter $\sigma_{\log \lambda}=0.228$, which we compare to the scatter in $\lambda^\prime$ allowed by our \rl\ relation below. 

The parameters $f_c(c)$ and $f_R(\lambda^\prime,m_d,c)$ in eq.~\ref{galmodr} account for the difference in total energy $E$ between an NFW and isothermal halo and the adiabatic contraction of the disk in the halo potential, respectively, where $c$ is the halo concentration. We use the exact expression for $f_c$ in eq.~23 of MMW98, and the approximate expression in their eq.~32 for $f_R$. Finally, we compute $c$ as a function of $L_d$ for our sample using the best-fitting linear relation between $c$ and $M_{200}$ for relaxed, {\it WMAP5} halos in \citet{maccio08} assuming $W = V_{200}/2$ \citep{dutton10}. 
 
Given the different measures of disk size presented in this paper, the definition of $R_d$ in eq.~\ref{galmodr} merits discussion. 
The model in eq.~\ref{galmodr} assumes that the (cold) baryons in disk galaxies are distributed in an exponential stellar disk with scale-length $R_d$. However, it is important to distinguish between $R_d$, the predicted scale-length of the stellar mass distribution, and $r_d$, the measured scale-length of the projected light distribution. For the variety of reasons discussed in \S\ref{intro}, extracting reliable deprojected scale-lengths from galaxy photometry is difficult. It is therefore not straightforward to relate the measured $r_d$ and the predicted $R_d$.

For a pure exponential disk with a fixed $\Upsilon_d$, $R_d$ is directly related to any given isophotal radius by the disk central surface brightness $\mu_0$.  If there is no scatter in $\mu_0$ at a given $L_d$, then the scatter in the distribution of isophotal radii is identical to the scatter in $R_d$. Moreover, if the median measured $r_d$ at a given $L_d$ is an unbiased estimator of $R_d$, then the ratio of $r_d$ to the median measured isophotal radius relates the latter directly to the predicted $R_d$. 

While the relationship between $R_d$ and isophotal radii described above clearly oversimplifies galactic structure, the basic scenario meshes with the properties of the SFI++. That the scatter in the $\log R_{23.5} - L$ relation is {\it 2.5 -- 4 times smaller} than that previously reported for the $r_d - L$ relation (see \S\ref{vrl}) evidences a disconnect between $r_d$ and $R_d$, and suggests that measured $R_{23.5}$ are better proxies for theoretical $R_d$ than measured scale-lengths $r_d$. We embrace this hypothesis in applying eq.~\ref{galmodr} to the nontemplate \rl\ relation presented in \S\ref{vrl}.

 \subsubsection{Application to the \rl\ Relation}
\begin{figure}
\epsscale{1.0}
\plotone{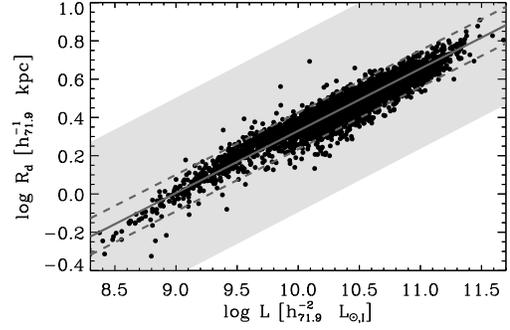}
\caption{\rl\ relation for the nontemplate subsample expressed in scale-length and luminosity units with $H_o = 71.9$ km s$^{-1}$ Mpc$^{-1}$. The solid line through the points shows eq.~\ref{galmodr} with parameters $\lambda^\prime = 0.042$, $m_d=0.054$, $\alpha=0.14$. The dashed lines show the scatter $2\epsilon_{obs} = 0.052 \,\log(h^{-1} \mathrm{kpc})$ about that relation allowed by the data, and the shaded region shows in comparison the scatter expected if the distribution of disk spin parameters $\lambda^\prime$  matched that of the spin parameters $\lambda$ of collisionless halos. \label{RLmod}}
\end{figure}

We now proceed to tune the parameters of eq.~\ref{galmodr} to match the \rl\ relation for the nontemplate subsample in Fig.~\ref{VRL_NT}. 
We convert $R_{23.5}$ to $R_d$ using the best-fitting linear relation in Fig.~\ref{rd_r235}, and convert to solar luminosities using $M_{\odot,I}=4.19$.  Fig.~\ref{RLmod} shows the \rl\ relation for the nontemplate subsample in these units.

 The solid line in Fig.~\ref{RLmod} shows eq.~\ref{galmodr} with our fiducial $\lambda^\prime = 0.042$ and $m_{d,0}=0.054$, with $\alpha=0.14$. This choice of parameters reproduces the slope and zero-point of the observed relation, but is not unique: \citet{dutton07} and \citet{gnedin07} demonstrate that a simultaneous fit of the scaling relations, their scatter and their residuals is required to break the degeneracies between parameters in eq.~\ref{galmodr}. 

Nonetheless, the implications of the scatter in our \rl\ relation for this model can be made clear by considering the maximum scatter in each of the parameters of eq.~\ref{galmodr} that is allowed by the data.  We model a constant scatter $\epsilon_{obs}$ (dashed lines in Fig.~\ref{RLmod}) with $\epsilon_{mes}=\epsilon_{RL}=0.04 \,\log (h^{-1} \mathrm{kpc})$ and $\epsilon_{int}=0.034\,\log (h^{-1} \mathrm{kpc})$ (see \S\ref{RL_budget}). The maximum scatter allowed in any one parameter of the model in eq.~\ref{galmodr} is therefore $\epsilon_{int}$. For simplicity, we assume that $f_R$ is scatter-free in this exercise. Eq.~\ref{galmodr} shows that the maximum scatters in $\lambda^\prime$ and $m_d$ allowed by the \rl\ relation are $\epsilon_{\log \lambda^\prime} = 0.034$ and $\epsilon_{\log md} = 0.102$, respectively. Since \citet{dutton07} find that the \rl\ relation limits the allowed range of these parameters even when simultaneous models to all the scaling relations are considered,  we consider these maximum scatters robust.  

This simple analysis suggests that the range of $\lambda^\prime$ and $m_d$ in spiral galaxies is significantly smaller than previously estimated: for example, our upper limit on the scatter in $\lambda^\prime$ is a factor of $\sim3$ smaller than that  found by \citet{dutton07}, because $\epsilon_{int}$ in our \rl\ relation scatter is significantly smaller than that of the one they model.  More importantly, $\epsilon_{\log \lambda^\prime}$ is a factor of 6.7 smaller than $\epsilon_{\log \lambda}$ found from cosmological simulations \citep{maccio08}: to illustrate, the shaded region in Fig.~\ref{RLmod} shows the expected scatter in the case where $\epsilon_{\log \lambda^\prime} = \epsilon_{\log \lambda}$. 

It is possible that the distribution of specific angular momentum $j_d/m_d$ in disk galaxies has less scatter than that of their host halos, perhaps due to a redistribution of angular momentum through bulge formation. This was explored by \citet{dutton07}, but their bulge formation scenario reduces the \rl\ scatter by at most $15\%$ for realistic bulge-to-disk ratios. It therefore seems unlikely that angular momentum redistribution is responsible for the narrow range of  $\lambda^\prime$ required by our model, particularly for the Sc-dominated SFI++. 

Another possibility is that disk galaxies form in a subset of halos with a distribution of spin parameters that is different from that in the halo population as a whole. Some studies suggest that this is the case: halos with a quiet merger history in the simulations of \citet{donghia04} have both lower $\bar{\lambda}$ and $\sigma_{\log \lambda}$ than found for all halos. This possibility has been previously discussed in the context of scaling relation models  \citep[e.g.][]{dejong00,pizagno05,dutton07,gnedin07}.  However, it is unclear whether this effect can explain the extremely small range of $\lambda$ implied by our \rl\ relation: once the scatter from other parameters such as $c$ and $\Upsilon_d$ are taken into account, the allowed scatter on $\lambda^\prime$ may well be an {\it order of magnitude} smaller than that predicted for $\lambda$ from cosmological expectations.

\section{Summary}

Of all the disk galaxy scaling relations, the \lv\ (i.e. Tully-Fisher) relation has by far received the most attention.  It has been used to compute distances to galaxies in the nearby universe and study the local velocity field, and to put constraints on models of galaxy formation, using data at different wavelengths and at both low and high redshift.   While the \rl\ and \rv\ relations can in theory serve similar purposes, the large scatter in previously published versions of these relations is prohibitive. 

With close to 5000 galaxies, the SFI++ is one of the largest available samples to study disk galaxy scaling relations.  The main advantage of the sample is the large range of carefully measured parameters available, and the well-studied uncertainties associated with them.  Homogeneous rotation velocities and $I-$band luminosities were published by S07, and the SFI++ \lv\ template relation studied in detail by M06. In addition to estimates of $V$ and $L$, measures of disk size $R$ are also available for a subset of the catalog. It is therefore possible to construct the full suite of $LRV$ scaling relations from the SFI++.

In this paper, we have presented the \lv, \rv\ and \rl\ relations for subsets of the template and nontemplate SFI++ samples. Because the SFI++ \lv\ relation has been extensively studied, we focus on the  \rv\ and \rl\ relations.  Contrary to previous studies, we adopt inclination-corrected isophotal radii rather than disk scale-lengths as a measure of the disk size $R$. While scale-length-based studies report an observed scatter for the \rl\ relation of $\epsilon_{obs}=0.15-0.20 \,\log (h^{-1} \mathrm{kpc})$ \citep[][P05, C07]{shen03,avila08}, we find $\epsilon_{obs}= 0.05\,\log (h^{-1} \mathrm{kpc})$.   With $\epsilon_{obs}=0.11 \log (h^{-1} \mathrm{kpc})$ and a correlation coefficient of $r=0.86$, our \rv\ relation is also significantly tighter than previously found. We demonstrate that the drastically tighter relations and lower scatters reported here stem from our use of isophotal radii, and argue that they are superior to measured scale-lengths because they are less susceptible to measurement errors and can be more reliably corrected for internal extinction. We examine and interpret the morphological dependence of the scaling relations as well as the properties of their residuals.

After carefully accounting for all known sources of measurement error and their covariances, we retrieve the amount of intrinsic scatter in each of the relations.  Our error budgets are consistent with constant intrinsic scatter $\epsilon_{int}$ at high $L$ and high $V$, with values  $\epsilon_{int} = 0.35\,$mag, $\epsilon_{int} = 0.084 \,\log (h^{-1} \mathrm{kpc})$ and $\epsilon_{int} =0.034 \,\log (h^{-1} \mathrm{kpc})$ for the \lv, \rv\ and \rl\ relations, respectively.  The scatter of the \lv\ and \rv\ relations however increases strongly at low $V$. While this likely reflects enhanced intrinsic scatter in $V$, the effect is difficult to quantify due to the competing contributions of peculiar velocities, sample incompleteness and measurement errors. On the other hand, the scatter in the \rl\ relation decreases slightly at low $L$, consistent with the behavior of an incompleteness bias.

We investigate the possible application of the \rv\ relation as a distance indicator. Comparing with the Cepheid variable distances available for 17 galaxies in the template subsample, we find that the  \rv\ relation returns distances with a median precision of 25\%. This is slightly inferior to the 15-20\% accuracy of \lv\ relation distances typically reported. We suggest that \rv\ distances are nonetheless accurate enough to provide a valid substitute when precise integrated magnitudes are not available. We find that the uncertainties on $H_0$ derived by matching the zero points of the \rv\ relations for the template subsample and Cepheid variable sample are comparable to those obtained using the \lv\ relation: in both cases, the uncertainties on the Cepheid distance zero point dominate.

To illustrate the potential of the SFI++ scaling relations to constrain the formation of Sc galaxies,  we apply a simple, MMW98-style galaxy formation model to the nontemplate \rl\ relation. Assuming that the intrinsic scatter in our measured isophotal radii equals that of the predicted scale-lengths of the stellar mass distribution, we find an upper limit on the scatter of the disk spin parameter that is 3 times smaller than previously reported, and 6.7 times smaller than the distribution of halo spin parameters predicted by cosmological simulations. As it is unlikely that angular momentum redistribution due to bulge formation can account for this difference, we suggest that the subset of halos in which Sc galaxies form has a much smaller spread in spin parameters than the broader halo population. A separate analysis will be required to quantify the full potential of the low-scatter SFI++ \rv\ and \rl\ relations for constraining galaxy formation in light of the improved size measurements presented here.

\acknowledgments
We thank Riccardo Giovanelli and Martha Haynes for advice on the contents of this paper and for kindly providing some unpublished parameters from the SFI++ catalog. We also thank Karen Masters and Christopher Springob for sharing their expertise about the details of the SFI++ catalog contents, and St\'ephane Courteau for suggestions that improved this paper. A.S. acknowledges support from the Swiss National Science Foundation (grant PP002-110576), and useful discussions with Kim-Vy Tran and Lea Giordano while working on this project at the University of Zurich. K. S. acknowledges support from the National Sciences and Engineering Research Council of Canada.



\appendix

\section{A. SFI++ Scaling Relation Parameters and their Uncertainties}
 \label{appendix1}

In \S\ref{meas_errors}, we summarize the derivation of disk luminosities and rotation velocities in the SFI++, and give an overview of our approach for computing homogeneous disk sizes for the sample studied here. In this appendix, we present the mathematical details of these computations. 

The standard error propagation formula for a function $f(x,y)$ is given by:
\begin{equation}
\epsilon_{f}^2 =   \left( \frac{\delta\,f}{\delta\,x} \right)^2\epsilon_x^2 +  \left( \frac{\delta\,f}{\delta\,y}  \right)^2 \epsilon_y^2 + 2  \left( \frac{\delta\,f}{\delta\,x}   \right) \left(\frac{\delta\,f}{\delta\,y}  \right)\epsilon_{xy} =  \delta_x^2\epsilon_x^2 + \delta_y^2\epsilon_y^2 + 2\delta_{x}\delta_y\epsilon_{xy}\,\,\,,
\label{covar}
\end{equation}
and can be expanded in the expected manner for functions of larger numbers of variables. In eq.~\ref{covar},  $\epsilon_x$ and $\epsilon_y$ are the uncertainties on the parameters $x$ and $y$, respectively, and $\epsilon_{xy}$ is the covariance between them.  We take covariances into account in the derivation of the SFI++ error budgets in Appendix~B.

\subsection{SFI++ Disk Luminosities: Absolute $I$-Band Magnitudes \label{ap_mcor}}
Disk luminosities in the SFI++ are expressed in terms of absolute $I$-band magnitudes. As explained in S07, the SFI++ absolute magnitude derivation  draws from the work of \citet{giovanelli97}, \citet{haynes99} and M06; for clarity, we use the notation of S07 where possible. The absolute magnitudes $M_{I}$ are given by: 
\begin{equation}
M_{I}=m_{obs}-A_I+\Delta M + k_I - 5\log \left [\frac{cz}{100 h} (1+z)\right ] -25,
\label{mcor}
\end{equation}
where $m_{obs}$ is the observed $I$-band magnitude extrapolated to 8 disk scale-lengths \citep{haynes99}, $A_I$ is the Galactic extinction term from the COBE/DIRBE maps of \citet{schlegel98}, and $k_I$ is a type- and redshift-dependent k-correction \citep{han92}:
\begin{equation}
k_I = (0.1658T - 0.5876)z\,.
\label{ki}
\end{equation}
We assign distances to the sample galaxies as described in \S\ref{distances}. 
The luminosity distance factor $(1+z)$ in the 5$^{\rm{th}}$ term on the right-hand side (RHS) of eq.~\ref{mcor}, not included in earlier SFI++ compilations, accounts for cosmological dimming.

The $\Delta M$ term in eq.~\ref{mcor} corrects for internal extinction, as first derived by \citet{giovanelli94}:
\begin{equation}
\Delta M = \gamma \log (1-e),
\end{equation}
where $e=1-b/a$ is the seeing-corrected ellipticity of the disk \citep{giovanelli97,haynes99} and $\gamma$ is a magnitude-dependent factor of order unity \citep[][S07]{giovanelli95}:
\begin{eqnarray}
 \nonumber
 \gamma & = & 0.5 \;\;\; \;\; \;\;\;\;\;\;\;\;\;\;\;\;\;\;\;\;\;\;\;\;\;\;\;\;\;\;\;\;\; \mathrm{for} \; M_{I}>-19.1\\
 \nonumber
                 & = & 1 - 0.417(M_I + 20.3) \;\;\;\;\;\,  \mathrm{for} \; -20.3 < M_{I} < -19.1 \\
 \nonumber                
                 & = & 1.0  \;\;\;\;\;\;\;\;\;\;\;\;\;\;\;\;\;\;\;\;\;\;\;\;\;\;\;\;\;\;\;\;\;\;   \mathrm{for} \; -21.8 < M_{I} < -20.3 \\
 \nonumber                
                 & = & 1.35 - 0.35(M_I + 22.8) \;\; \,\mathrm{for} \; -22.7 < M_{I} < -21.8 \\
                 & = & 1.3 \;\;\; \;\; \;\;\;\;\;\;\;\;\;\;\;\;\;\;\;\;\;\;\;\;\;\;\;\;\;\;\;\;\;   \mathrm{for} \; M_{I} < -22.7
\label{gammaval}
 \end{eqnarray}

 We note that there are differences of $\sim 0.1\,$mag between the $M_{I}$  that we compute using eq.~\ref{mcor} and the corresponding values listed in tables~2~and~4 of S07 (specifically, the values listed in the erratum to that paper). The two main reasons for this difference are (1) $M_{I}$ listed in S07 have the morphological correction of M06 applied. By contrast, we only correct $M_I$ for morphology in the \lv\ relation, in the manner described in \S\ref{morphology}; and (2) M06 and S07 actually use a {\it velocity width}-dependent value of $\gamma$ in computing their $M_I$, rather than their published relation in eq.~\ref{gammaval} (K.\ L.\ Masters, priv.\ comm.). We have verified that these small differences do not affect the results presented here.
 
 The measurement uncertainties $\epsilon_{M}$ on $M_{I}$ are computed as in \citet{giovanelli97}.  Contributions from $cz$ ($<1\%$) and $k_I$ ($<0.01$ mag) are small compared to other factors, and are therefore ignored. Assuming that the uncertainties $\epsilon_{m}$ on $m_{obs}$, $\epsilon_{A}$ on $A_I$, $\epsilon_{\gamma}$ on $\gamma$, and $\epsilon_e$ on $e$ are uncorrelated and using eq.~\ref{covar}, $\epsilon_{M}$ is given by:
\begin{equation}
\epsilon_{M}^2 = \epsilon_{m}^2 + \epsilon_A^2 + \delta_{\gamma}^2 \epsilon_{\gamma}^2 + \delta_{eM}^2 \epsilon_e^2, 
\label{emcor}
\end{equation}
where 
\begin{equation}
\delta_{\gamma}=\log(1-e),
\label{eq_deltagamma}
\end{equation}
\begin{equation}
\delta_{eM}=\frac{-\gamma}{\ln (10) (1-e)},
\label{eq_deltaeM}
\end{equation}
In the SFI++, $\epsilon_{mobs}$ and $\epsilon_e$ are computed using the method of \citet{haynes99b}, $\epsilon_{\gamma}=0.15\gamma$ \citep{giovanelli97} and $\epsilon_A = 0.2A_I$ (C.\ M.\ Springob, priv. comm.).  

\subsection{SFI++ Disk Rotation Velocities: Velocity Widths \label{ap_wcor}}
  Disk rotation velocities in the SFI++ are expressed in terms of velocity widths. As explained in S07, these widths are homogeneously derived from either single-dish HI measurements or ORCs, with a preference for the former when both are available for a given galaxy. The corrected velocity widths in the SFI++ are given by:
\begin{equation}
W = \frac{Q}{\sin{i}},
\label{W}
\end{equation}
where
\begin{equation}
Q= \Bigg \{ \begin{array}{ll} 
   \frac{W_{obs,21}-\Delta_s}{1+z} - \Delta_t  & \mbox{for HI widths},\\
   \frac{W_{obs,ORC}}{1+z} \left [ \frac{W_{21}}{W_{ORC}} \right ] & \mbox{for ORC widths.}
   \end{array}
   \label{eqQ}
\end{equation}

The HI width measurement technique and corrections are given in \citet{springob05}. Briefly, $W_{21,obs}$ in eq.~\ref{eqQ} is the observed width and $\Delta_t = 6.5\,$\kms\ is the correction for turbulent motions.  The instrumental broadening is $\Delta_s = 2\Delta{v}\lambda$, where $\Delta{v}$ is the spectrometer channel separation in \kms\ and the parameter $\lambda$ depends on the spectrum quality and smoothing (see table~2 of \citealt{springob05}). 

The ORC width derivation is given in \citet{catinella05} and \citet{catinella07}: as explained in those papers, the  measured width $W_{ORC,obs}$ in eq.~\ref{eqQ} is the parametric {\it Polyex} model fit \citep{giovanelli02} at the radius $r_{opt}$ containing 83\% of the total $I$-band light. Since the HI disks of spirals typically extend to twice this value, a correction factor $W_{21}/W_{ORC}$ is applied to homogenize the HI and ORC widths:
\begin{equation}
\frac{W_{21}}{W_{ORC}}= \Bigg \{ \begin{array}{ll}
    0.899 + 0.188 r_{max}/r_{opt} & \mbox{for rising ORCs},\\
    1.075 - 0.013 r_{max}/r_{opt} & \mbox{for flat ORCs},
    \end{array}
    \label{cor_ORC}
\end{equation}
where $r_{max}$ is the measured extent of the ORC.

 The disk inclination $i$ is computed from $e$ using:
 \begin{equation}
(\cos i)^2 = \frac{(1-e)^2-q_0^2}{1-q_0^2}, 
\label{cosi}
\end{equation}
where the intrinsic axial ratio $q_0=0.13$ for galaxies of types Sbc and later, and $q_0=0.20$ for earlier types \citep[][S07]{giovanelli94}.
Substituting for $\sin{i}$ using eq.~\ref{cosi} and adopting logarithmic units, eq.~\ref{W} can be rewritten as:
\begin{equation}
\log W = \log Q + 0.5\log (1-q_0^2) - 0.5\log[1-(1-e)^2].
\label{logw}
\end{equation}
Assuming that the measurement uncertainties on individual parameters are uncorrelated and applying eq.~\ref{covar}, the net measurement uncertainty $\epsilon_W$ on $\log W$ is therefore:
\begin{equation}
\epsilon_W^2 \equiv \epsilon_{\log W}^2 = \delta_Q^2 \epsilon_Q^2 + \delta_{q}^2 \epsilon_{q}^2 + \delta_{eW}^2 \epsilon_e^2,
\label{elogW}
\end{equation}
where
\begin{equation}
\delta_Q=\frac{1}{\ln(10) Q},
\label{eq_deltaQ}
\end{equation}
\begin{equation}
\delta_{q} = \frac{-q_0}{\ln(10) (1-q_0^2)},
\label{eq_deltaq0}
\end{equation}
\begin{equation}
\delta_{eW}=\frac{-(1-e)}{\ln(10)(1-[1-e]^2)}.
\label{eq_deltaeW}
\end{equation}
The uncertainty $\epsilon_Q$ in $Q$ can be derived from eqs.~\ref{covar} and \ref{eqQ} assuming uncorrelated errors:
\begin{equation}
\epsilon_Q= \Bigg \{ \begin{array}{ll} 
   \sqrt{\frac{\epsilon_{Wobs}^2+\epsilon_{\Delta s}^2}{(1+z)^2} + \epsilon_{\Delta t}^2}  & \mbox{for HI widths}\\
   \frac{\epsilon_{Wobs}}{(1+z)}\left[\frac{W_{21}}{W_{ORC}}\right]  & \mbox{for ORC widths,}
   \end{array}
\end{equation}
where $\epsilon_{Wobs}$ is the width measurement uncertainty, and the uncertainty on  $W_{21}/W_{ORC}$ is assumed to be negligible. In the SFI++, $\epsilon_{\Delta s}=0.25 \Delta_s$, $\epsilon_{\Delta t}=0.25 \Delta_t$, and $\epsilon_{q}=0.15q_0$ \citep{giovanelli97}. 

The values of $\log W$ and $\epsilon_W$ computed here are identical to parameters $\log(W_{TF})$ and $\epsilon_w$ published in tables~2~and~4 of S07.

\subsection{SFI++ Disk Sizes: Isophotal Radii \label{ap_rcor}}

 As explained in \S\ref{meas_rcor}, we adopt the radius of the disk corresponding to the $\mu_I=23.5$ mag arcsec$^{-2}$ isophote as the SFI++ disk size:
\begin{equation}
r_{23.5}^o = r_{23.5} \left [ (1-e)^{-\beta} + \Delta R \right ]^{-1},
\label{roeq}
\end{equation}
where the terms in brackets on the right-hand side correct for internal extinction in the disk following the prescription of \citet{giovanelli95}. In eq.~\ref{roeq}, $\Delta R$ is a small perturbative adjustment to account for photometric profile variations in galaxies with a given $M_{I}$:
\begin{equation}
\Delta R=- \ln(10) \log (1-e) (1-e)^{\alpha-2\beta} \left ( \frac{r_d}{r_{23.5}} - \langle \frac{r_d}{r_{23.5}} \rangle \right ),
\label{eq_DeltaR}
\end{equation}
(c.f.\ eq.~7 of \citealt{giovanelli95}) where $\alpha$ and $\beta$ are luminosity-dependent factors shown in their fig.~7. We use the best-fitting linear relations:
\begin{equation}
\alpha = 0.041 + 1.0 M_{I} \,\,\,,
\end{equation}
\begin{equation}
\beta = 0.031 + 0.863 M_{I}\,\,\,.
\end{equation}
with $M_I$ from eq.~\ref{mcor}. In eq.~\ref{eq_DeltaR}, $r_{d}$ is the measured disk scale-length and $\langle r_{d}/r_{23.5} \rangle$ is the average ratio of disk scale-lengths to isophotal radii as a function of $M_I$ for our sample, given by eq.~\ref{rdeq}.  

The parameter $r_{23.5}$ in eqs.~\ref{roeq}~and~\ref{eq_DeltaR} is the measured isophotal radius $r_{obs,23.5}$ corrected for galactic extinction $A_I$,  a $k$-correction $k_I$ and cosmological surface brightness dimming per unit frequency interval:
\begin{equation}
r_{23.5}= r_{obs,23.5} + \frac{\ln(10)r_d}{2.5} [A_I - k_I + 2.5\log(1+z)^3].
\label{r235}
\end{equation}
 We use the same values of $A_I$, $k_I$ and $z$ as adopted to compute $M_{I}$ (eq.~\ref{mcor}).

Expressing eq.~\ref{roeq} in physical, logarithmic units, we obtain:
\begin{equation}
\log{R_{23.5}}= \log(r_{23.5}^o) + \log \left( \frac{cz}{100h(1+z)} \right) - \log B,
\label{rcor}
\end{equation} 
where we have adopted the angular size distance and $B$ converts angular size units to radians. 
We note that the cosmological surface brightness dimming term in eq.~\ref{r235} partly offsets the cosmological stretching implied by the angular size distance in eq.~\ref{rcor}.

 We assume that the measurement errors on the individual parameters used to derive $\log R_{23.5}$ are uncorrelated, and following \citet{giovanelli97} we neglect the contributions from $cz$ and $k_I$. We note that because the perturbative correction term $\Delta R$ is small compared to the measurement errors $\epsilon_{r}$ on $r_{obs,23.5}$ for most galaxies in the sample, we assume that the uncertainty on this correction simply scales with its amplitude.  The measurement errors on $\log R_{23.5}$ are therefore:
\begin{equation}
\epsilon_{R}^2 \equiv \epsilon_{logR23.5}^2 = \delta_{r}^2 \epsilon_{r}^2 +\delta_{rd}^2 \epsilon_{rd}^2 + \delta_{AR}^2\epsilon_A^2 + \delta_{\beta}^2 \epsilon_{\beta}^2 +\delta_{\Delta R}^2 \epsilon_{\Delta R}^2+ \delta_{eR}^2 \epsilon_{e}^2 , 
\label{ercor}
\end{equation}
where 
\begin{equation}
\delta_{r}=\frac{1}{ \ln(10)r_{23.5}},
\label{eq_deltarobs}
\end{equation}
\begin{equation}
\delta_{rd}=\frac{A_I - k_I +2.5\log(1+z)^3}{2.5 r_{23.5} },
\label{eq_deltard}
\end{equation}
\begin{equation}
\delta_{AR} = \frac{r_d}{2.5r_{23.5}},
\label{eq_deltaar}
\end{equation}
\begin{equation}
\delta_{\beta}=\frac{r_{23.5}^o}{r_{23.5}} \left [ \frac{\log(1-e)}{(1-e)^{\beta}} \right ],
\label{eq_deltabeta}
\end{equation}
\begin{equation}
\delta_{\Delta R}=-\frac{1}{\ln (10)} \left ( \frac{r_{23.5}^o}{r_{23.5}} \right ),
\label{eq_deltaDeltaR}
\end{equation}
\begin{equation}
\delta_{eR}=-\frac{r_{23.5}^o}{r_{23.5}} \left ( \frac{\beta}{\ln(10) (1-e)^{1+\beta}} \right ).
\label{eq_deltaeR}
\end{equation}

 Following the approach of \citet{giovanelli97}, we adopt uncertainties of $\epsilon_{\beta}=0.15\beta$ and $\epsilon_{\Delta R}=0.15\Delta R$. Individual measurement uncertainties $\epsilon_{r}$ on $r_{obs,23.5}$ and $\epsilon_{rd}$ on $r_d$ are not available for SFI++ galaxies. We therefore adopt appropriate characteristic values $\epsilon_{r}=0.05r_{obs,23.5}$ and $\epsilon_{rd}=0.15r_{d}$ (M.\ P.\ Haynes, priv.\ comm.).

\section{B. Measurement Uncertainties in the SFI++ Scaling Relations }
\label{appendix2}

 Below, we derive relations describing the measurement error contributions $\epsilon_{mes}$ for the scaling relations constructed from $R$, $L$ and $V$ (see eq.~\ref{scat_obs}), accounting for correlations between the measurement errors in these quantities. We also derive the expected contribution to the intrinsic scatter of each relation from galaxy peculiar velocities.

\subsection{The SFI++ \lv\ Relation}
The \lv\ relation in the SFI++ is expressed as a linear correlation between $L = M_{I}$ and $V = \log W$ \citep[][M06, S07]{giovanelli97}:
\begin{equation}
M_{I} = a_{LV} + b_{LV} \log W \;,
\end{equation}
where $a_{LV}$ and $b_{LV}$ are the best-fitting zero point and slope. The net contribution of measurement uncertainties to this scatter can be derived from the expressions for $M_{I}$ (eq.~\ref{mcor}) and $\log{W}$ (eq.~\ref{logw}):

\begin{equation}
 \epsilon_{LV}^2\equiv \epsilon_{mes}^2 =\epsilon_{m}^2 + \epsilon_A^2+\delta_{\gamma}^2\epsilon_{\gamma}^2 + (b_{LV} \delta_Q)^2 \epsilon_Q^2 + (b_{LV}\delta_{q})^2\epsilon_{q}^2+\delta_{eLV}^2\epsilon_e^2 + \epsilon_{\Delta LV}^2 \;,
\label{epsilonLV}
\end{equation}
where $\delta_{\gamma}$, $\delta_Q$ and $\delta_{q}$ are given in equations \ref{eq_deltagamma}, \ref{eq_deltaQ} and \ref{eq_deltaq0}, respectively,  and
\begin{equation}
\delta_{eLV}=\delta_{eM}-b_{LV}\delta_{eW} \;,
\end{equation}
where $\delta_{eM}$ and $\delta_{eW}$ are given in eqs.~\ref{eq_deltaeM} and \ref{eq_deltaeW} \citep[see also][M06]{giovanelli97}.  The parameter $\epsilon_{\Delta LV}$ is the uncertainty in the morphological type correction $\Delta_{LV}$ computed in eq.~\ref{deltaLV}, which we take as $\epsilon_{\Delta LV} = 0.15 \Delta_{LV}$. We note that given the number of galaxies in the SFI++, the statistical uncertainties on $a_{LV}$ and $b_{LV}$ are small compared to parameter measurement uncertainties, and their contribution to the error budget can be neglected.  

It is clear that since $\log W$ is a distance-independent quantity, peculiar velocities $V_{pec}$ introduce scatter  in the \lv\ relation only in $M_I$.  Considering the 5th term on the RHS of eq.~\ref{mcor}, a peculiar velocity $V_{pec}$ contributes a scatter
\begin{equation}
\epsilon_{LV,Vpec} = 5 \log ( 1 + V_{pec}/cz ) \;,
\label{LV_Vpec}
\end{equation}

\subsection{The SFI++ \rv\ Relation}
We express the SFI++ \rv\ relation as a linear relation between $R = \log R_{23.5}$ (eq.~\ref{rcor}) and $V = \log W$ (eq.~\ref{logw}):
\begin{equation}
\log R_{23.5} = a_{RV} + b_{RV} \log W \;,
\end{equation}
where $a_{RV}$ and $b_{RV}$ are the best-fitting zero-point and slope. The total measurement error contribution $\epsilon_{RV}$ to the scatter in the \rv\ relation is therefore:

\begin{equation}
 \epsilon_{RV}^2 \equiv \epsilon_{mes}^2 =\delta_{r}^2 \epsilon_{r}^2 +\delta_{rd}^2 \epsilon_{rd}^2 + \delta_{AR}^2\epsilon_A^2+ \delta_{\beta}^2 \epsilon_{\beta}^2 +\delta_{\Delta R}^2 \epsilon_{\Delta R}^2+(b_{RV}\delta_Q)^2\epsilon_Q^2 +(b_{RV}\delta_{q})^2\epsilon_{q}^2  +\delta_{eRV}^2 \epsilon_{e}^2 + \epsilon_{\Delta RV}^2 \;,
\label{epsilonRV}
\end{equation}
where
\begin{equation}
\delta_{eRV} = \delta_{eR}-b_{RV}\delta_{eW} \;.
\label{deltaeRV}
\end{equation}
In eq.~\ref{epsilonRV}, $\delta_{r}$, $\delta_{rd}$, $\delta_{AR}$, $\delta_{\beta}$, $\delta_{\Delta R}$, $\delta_{Q}$ and $\delta_q$ are given by eqs.~\ref{eq_deltarobs}, \ref{eq_deltard}, \ref{eq_deltaar}, \ref{eq_deltabeta},  \ref{eq_deltaDeltaR}, \ref{eq_deltaQ} and \ref{eq_deltaq0}, respectively.  In eq.~\ref{deltaeRV}, $\delta_{eR}$ and $\delta_{eW}$ are given by eqs. \ref{eq_deltaeR} and \ref{eq_deltaeW}. The parameter $\epsilon_{\Delta RV}$ accounts for the uncertainty in the morphological correction of eq.~\ref{deltaRV}, which we take to be $\epsilon_{\Delta RV} = 0.15\Delta_{RV}$. 

As in the \lv\ relation, peculiar velocities introduce scatter in the \rv\ relation only along the $\log R_{23.5}$ axis because $\log W$ is a distance-independent quantity. The second term on the RHS of eq.~\ref{rcor} shows that the scatter along this axis for non-zero $V_{pec}$ is:
\begin{equation}
\epsilon_{RV,Vpec} = - \log ( 1 + V_{pec}/cz ) \;,
\label{RV_Vpec}
\end{equation}

\subsection{The SFI++ \rl\ relation}
We express the SFI++ \rl\ relation as a linear relation between $R = \log R_{23.5}$ and $L = M_{I}$: 
\begin{equation}
\log R_{23.5} = a_{RL} + b_{RL}M_{I} \;.
\end{equation}
 The total measurement error contribution to the scatter in the \rl\ relation is therefore:
\begin{equation}
\epsilon_{RL}^2 \equiv \epsilon_{mes}^2=\delta_{r}^2 \epsilon_{r}^2 + \delta_{rd}^2 \epsilon_{rd}^2 + \delta_{\beta}^2 \epsilon_{\beta}^2 +\delta_{\Delta R}^2 \epsilon_{\Delta R}^2+b_{RL}^2\epsilon_{m}^2 +(b_{RL}\delta_{\gamma})^2\epsilon_{\gamma}^2  + \delta_{ARL}^2\epsilon_A^2+\delta_{eRL}^2 \epsilon_{e}^2 + \epsilon_{\Delta RL}^2 \;,
\label{epsilonRL}
\end{equation}
where
\begin{equation}
\delta_{ARL} = \delta_{AR} - b_{RL} \;,      \;\;\;\;\;\;\;\;\;\; \delta_{eRL} = \delta_{eR}-b_{RL}\delta_{eM} \; .
\label{deltaARL}
\end{equation}

In eq.~\ref{epsilonRL}, $\delta_{r}$, $\delta_{rd}$, $\delta_{\beta}$, $\delta_{\Delta R}$ and $\delta_\gamma$ are given by eqs.~\ref{eq_deltarobs}, \ref{eq_deltard}, \ref{eq_deltabeta}, \ref{eq_deltaDeltaR} and \ref{eq_deltagamma}, respectively, while $\delta_{eR}$ and $\delta_{eM}$ in eq.~\ref{deltaARL} are given by eqs.~\ref{eq_deltaeR} and \ref{eq_deltaeM}.  The parameter $\epsilon_{\Delta RL}$ accounts for the uncertainty in the morphological correction of eq.~\ref{deltaRL}, which we adopt as $\epsilon_{\Delta RL} = 0.15 \Delta_{RL}$. 

Unlike the \rv\ and \lv\ relations, peculiar velocities introduce scatter in {\it both} variables of the \rl\ relation, since $\log R_{23.5}$ and $M_I$ depend on distance. Eqs.~\ref{RV_Vpec}~and~\ref{LV_Vpec} show that a positive peculiar velocity $V_{pec}$ simultaneously scatters points towards fainter $L$ and smaller $R$.  The amplitude of the scatter is:
\begin{equation}
\epsilon_{RL,Vpec} = \sqrt { [5 \log ( 1 + V_{pec}/cz )]^2 +  [\log ( 1 + V_{pec}/cz )]^2 } = \sqrt{26} \log ( 1 + V_{pec}/cz ) \;.
\end{equation}
This scatter vector has a constant slope in the $L $--$R$ plane:
\begin{equation}
S_{RL,Vpec} = \frac{- \log (1 + V_{pec}/cz )}{ 5 \log ( 1 + V_{pec}/cz )} = -0.2 \; ,
\end{equation}
this value is similar to the best-fitting slope $\beta = -0.16$ of the measured \rl\ relations (Table~\ref{templates}).  Peculiar velocities therefore scatter points almost exactly {\it along} the \rl\ relation, and do not contribute significantly to the intrinsic scatter in the relation.

\end{document}